\PassOptionsToPackage{unicode}{hyperref}
\PassOptionsToPackage{hyphens}{url}
\PassOptionsToPackage{dvipsnames,svgnames,x11names}{xcolor}
\documentclass[12pt]{article}

\usepackage{amsmath,amssymb}
\usepackage{amsthm}
\usepackage{comment}
\usepackage{multirow}
\usepackage{multicol}
\usepackage{enumitem}

\usepackage{iftex}
\ifPDFTeX
  \usepackage[T1]{fontenc}
  \usepackage[utf8]{inputenc}
  \usepackage{textcomp} 
\else 
  \usepackage{unicode-math}
  \defaultfontfeatures{Scale=MatchLowercase}
  \defaultfontfeatures[\rmfamily]{Ligatures=TeX,Scale=1}
\fi
\usepackage{lmodern}
\ifPDFTeX\else  
\fi
\IfFileExists{upquote.sty}{\usepackage{upquote}}{}
\IfFileExists{microtype.sty}{
  \usepackage[]{microtype}
  \UseMicrotypeSet[protrusion]{basicmath} 
}{}
\makeatother
\usepackage{xcolor}
\makeatletter
\ifx\paragraph\undefined\else
  \let\oldparagraph\paragraph
  \renewcommand{\paragraph}{
    \@ifstar
      \xxxParagraphStar
      \xxxParagraphNoStar
  }
  \newcommand{\xxxParagraphStar}[1]{\oldparagraph*{#1}\mbox{}}
  \newcommand{\xxxParagraphNoStar}[1]{\oldparagraph{#1}\mbox{}}
\fi
\ifx\subparagraph\undefined\else
  \let\oldsubparagraph\subparagraph
  \renewcommand{\subparagraph}{
    \@ifstar
      \xxxSubParagraphStar
      \xxxSubParagraphNoStar
  }
  \newcommand{\xxxSubParagraphStar}[1]{\oldsubparagraph*{#1}\mbox{}}
  \newcommand{\xxxSubParagraphNoStar}[1]{\oldsubparagraph{#1}\mbox{}}
\fi
\makeatother

\usepackage{longtable,booktabs,array}
\usepackage{calc} 
\usepackage{etoolbox}
\makeatletter
\patchcmd\longtable{\par}{\if@noskipsec\mbox{}\fi\par}{}{}
\makeatother
\IfFileExists{footnotehyper.sty}{\usepackage{footnotehyper}}{\usepackage{footnote}}
\makesavenoteenv{longtable}
\usepackage{graphicx}
\makeatletter
\def\maxwidth{\ifdim\Gin@nat@width>\linewidth\linewidth\else\Gin@nat@width\fi}
\def\maxheight{\ifdim\Gin@nat@height>\textheight\textheight\else\Gin@nat@height\fi}
\makeatother
\setkeys{Gin}{width=\maxwidth,height=\maxheight,keepaspectratio}
\makeatletter
\def\fps@figure{htbp}
\makeatother

\makeatletter
\@ifpackageloaded{caption}{}{\usepackage{caption}}
\AtBeginDocument{%
\ifdefined\contentsname
  \renewcommand*\contentsname{Table of contents}
\else
  \newcommand\contentsname{Table of contents}
\fi
\ifdefined\listfigurename
  \renewcommand*\listfigurename{List of Figures}
\else
  \newcommand\listfigurename{List of Figures}
\fi
\ifdefined\listtablename
  \renewcommand*\listtablename{List of Tables}
\else
  \newcommand\listtablename{List of Tables}
\fi
\ifdefined\figurename
  \renewcommand*\figurename{Figure}
\else
  \newcommand\figurename{Figure}
\fi
\ifdefined\tablename
  \renewcommand*\tablename{Table}
\else
  \newcommand\tablename{Table}
\fi
}
\@ifpackageloaded{float}{}{\usepackage{float}}
\floatstyle{ruled}
\@ifundefined{c@chapter}{\newfloat{codelisting}{h}{lop}}{\newfloat{codelisting}{h}{lop}[chapter]}
\floatname{codelisting}{Listing}

\makeatother
\makeatletter
\@ifpackageloaded{caption}{}{\usepackage{caption}}
\@ifpackageloaded{subcaption}{}{\usepackage{subcaption}}
\makeatother

\ifLuaTeX
  \usepackage{selnolig}  
\fi
\usepackage[]{natbib}
\usepackage{bookmark}

\IfFileExists{xurl.sty}{\usepackage{xurl}}{} 
\hypersetup{
  pdftitle={Title},
  pdfauthor={Author 1; Author 2},
  pdfkeywords={3 to 6 keywords, that do not appear in the title},
  colorlinks=true,
  linkcolor={blue},
  filecolor={Maroon},
  citecolor={Blue},
  urlcolor={Blue},
  pdfcreator={LaTeX via pandoc}}

\newcommand{\anon}{1}

\theoremstyle{definition}

\newtheorem{proposition}{Proposition}[section]
\newtheorem{theorem}{Theorem}[section]

\newtheorem{lemma}[theorem]{Lemma}

\theoremstyle{definition}

\newtheorem{assumption}{Assumption}

\newcommand{\utwi}[1]{{\boldsymbol{#1} }}

\newcommand{\bff}{{\utwi{f}}}

\newcommand{\bs}{{\utwi{s}}}

\newcommand{\bu}{{\utwi{u}}}
\newcommand{\bv}{{\utwi{v}}}
\newcommand{\bw}{{\utwi{w}}}
\newcommand{\bx}{{\utwi{x}}}
\newcommand{\by}{{\utwi{y}}}
\newcommand{\bz}{{\utwi{z}}}
\newcommand{\bA}{{\utwi{A}}}

\newcommand{\bE}{{\utwi{E}}}
\newcommand{\bF}{{\utwi{F}}}

\newcommand{\bM}{{\utwi{M}}}

\newcommand{\bU}{{\utwi{U}}}

\newcommand{\bX}{{\utwi{X}}}
\newcommand{\bY}{{\utwi{Y}}}

\newcommand{\bSigma}{{\utwi{\Sigma}}}

\newcommand{\bphi}{{\utwi{\phi}}}

\newcommand{\btheta}{{\utwi{\theta}}}

\newcommand{\cE}{{\cal E}}

\newcommand{\cL}{{\cal L}}

\newcommand{\cX}{{\cal X}}

\let\oldtop\top 
\renewcommand\top{{^\oldtop}}

\newcommand{\ones}{\boldsymbol{1}}

\def\E{\mathbb{E}}
\def\P{\mathbb{P}}
\def\RR{\mathbb{R}}
\def\R{\mathbb{R}}

\begin{document}

\def\spacingset#1{\renewcommand{\baselinestretch}%
{#1}\small\normalsize} \spacingset{1}


\if1\anon
{
  \title{\bf Threshold Tensor Factor Model in CP Form}
  \author{Stevenson Bolivar\hspace{.2cm}\\
    Department of Statistics, Rutgers University\\
    and \\
    Rong Chen \thanks{Rong Chen's research was supported in part by National Science Foundation grants DMS-2319260 and DMS-2413858.
    }\\
    Department of Statistics, Rutgers University\\
    and \\
    Yuefeng Han \thanks{Yuefeng Han's research was supported in part by National Science Foundation grant DMS-2412578.}\\
    Department of Applied and Computational Mathematics and Statistics, \\
    University of Notre Dame
    }
  \maketitle
} \fi

\if0\anon
{
  \bigskip
  \bigskip
  \bigskip
  \begin{center}
    {\LARGE\bf Threshold Tensor Factor Model in CP Form}
\end{center}
  \medskip
} \fi

\bigskip
\begin{abstract}
This paper proposes a new Threshold Tensor Factor Model in Canonical Polyadic (CP) form for tensor time series. By integrating a thresholding autoregressive structure for the latent factor process into the tensor factor model in CP form, the model captures regime-switching dynamics in the latent factor processes while retaining the parsimony and interpretability of low-rank tensor representations. We develop estimation procedures for the model and establish the theoretical properties of the resulting estimators. Numerical experiments and a real-data application illustrate the practical performance and usefulness of the proposed framework.

\end{abstract}

\noindent%
\vfill

\newpage
\spacingset{1.7} 

\section{Introduction}
    Tensor time series data, denoted as $\mathcal{X}_t \in \mathbb{R}^{d_1 \times d_2 \times \cdots \times d_K}$ for $t = 1, \ldots, T$, is a collection of multidimensional arrays observed sequentially over time. 
Such tensor time series arise in various scientific and financial applications, including multi-category product import-export volume among a group of countries over time \citep{ChenYangZhang2022}, quarterly economic indices of different countries \citep{ChenChenTsay2019}, returns of portfolio constructed by size, book to market ratio, and momentum \citep{LiXiao2021}, multivariate spatial-temporal data \citep{chen2020modeling,barigozzi2025general}, and many others. 

\cite{Bolivar2025Review} provides a review of recent developments in autoregressive and factor modeling for such data. For example, \cite{ChenXiaoYang2021}, \cite{han2023rr} and \cite{li2024cointegrated} 
extend the autoregressive model to matrix-valued time series (MAR), while \cite{LiXiao2021} and \cite{WangZhengLi2024} extend it further to tensor time series. Another widely used approach in high-dimensional data is dimension reduction through factor models. The two main decompositions for tensor data structures are Tucker decomposition and CP decomposition \citep{KodaBader2009}. Mimicking Tucker decomposition, \cite{WhangLiuChen2019}, \cite{ChenYangZhang2022} and  \cite{ChenFan2023} introduced a tensor factor model in Tucker form, 
with factors in tensor form but of much smaller dimensions than the observed tensors. Other notable contributions in this area include \cite{ChenChenTsay2019,han2024tensor,han2022rank,chen2025diffusion}, and others.

In contrast, 
\cite{han2024cp} propose a tensor factor model in CP form (TFM-cp) that uses a set of univeriate factor processes that are uncorrelated, and rank-1 tensor factor loadings.
This approach mimics the CP decomposition. This model structure facilitates the analysis of underlying dynamics in high-dimensional tensor time series through a set of univariate uncorrelated latent factor processes, making it significantly more convenient for studying the dynamics of the time-series. The TFM-cp is specified as:
\begin{align}\label{TFM-cp}
\mathcal{X}_t = \sum_{j=1}^r \lambda_j f_{jt} \mathbf{u}_{j1} \circ \mathbf{u}_{j2} \circ \cdots \circ \mathbf{u}_{jK} + \mathcal{E}_t,
\end{align}
where each $f_{jt}$ is a one-dimensional latent factor, $\bu_{jk} \in \mathbb{R}^{d_k}$ are the loading vectors normalized to $\|\bu_{jk}\|_2=1$, $\cE_t$ is a zero mean idiosyncratic noise tensor with potentially weak cross-correlations among its elements and assumed to be uncorrelated with the latent factors, and $r$ is the rank of the decomposition, or equivalently the number of factors. Without loss of generality, it is assumed $\E f_{jt}^2\le C<\infty$, for all $1\le j\le r$. Then, the signal strengths are contained in $\lambda_j$. Here, $\{\bu_{jk}, 1\le i\le r\}$ are not necessarily orthogonal. \cite{chang2023modelling,han2024cp} showed that the model can be very useful in many applications.

We observe that the existing approaches mentioned above are all linear models. It has been widely acknowledged that in many applications linear time series models are not sufficient to capture various nonlinear phenominons encountered and there has been a large literature in study nonlinear time series models \citep{Tong90, tsay2018nonlinear,Gooijer2017NonLin_book}. 
Modern machine learning algorithms such as deep neural networks are also large nonlinear systems and have shown to be extremely powerful in providing accurate predictions \citep{zhou2025factor,luo2025supervised}. 
 
Motivated by observed regime-switching phenomenon in applications and encouraged by its simple form, regime-switching models, particularly the threshold models, are a class of powerful nonlinear time series models that have been studied extensively and have been shown to be a simple, elegant and useful parametric model 
\citep{Tong90, Chan&Tong86, Sims&Zha2006, tsay1989testing, ang2012regime,hansen2011threshold} in many applications ranging from economics, biology and ecology, signal processing, and many others. As an
attempt to move from the linear models to nonlinear ones for tensor time series, in this paper we propose a Threshold dynamic Tensor Factor Model in CP form (T-TFM-cp), which extends the standard TFM-cp model by incorporating a threshold mechanism into the latent factor dynamics. Specifically, we model the factors $f_{jt}$ using a general Threshold Autoregressive (TAR) model:
\begin{align} \label{tar0}
f_{jt} &= \text{TAR}(p_j, s_j, z_{jt}), \quad j = 1, \ldots, r.
\end{align} 
The TAR model allows the factor dynamics to switch between different regimes depending on the value of a threshold variable $z_{jt}$. 
This variable can be a function of observable variables (at time $t$) or lagged latent factors. The parameters $p_j$ and $s_j$ determine the specific form of the TAR model for factor $j$.

There is a vast literature on TAR models \citep{Tong90, Chan&Tong86, Sims&Zha2006, tsay1989testing, ang2012regime}. For matrix autoregressive models, recent work such as \cite{Bucci2024}, \cite{WuChan2024}, and \cite{YuLiZhangTong2024} develops new methods for modeling regime-specific changes in the coefficient matrices.
Related but conceptually distinct, \cite{LiuChen2020} introduce a threshold vector factor model, and \cite{LiuChen2022} study a threshold matrix factor model in Tucker form. In their framework, the loading matrices differ across regimes and regime switching is controlled by a threshold variable. The latent factors are assumed to be stationary but are not equipped with an explicit dynamic model, and therefore these approaches do not provide forecasting capability. 
In contrast, the proposed T-TFM-cp model assumes common loading structures across all regimes, while allowing the latent factor processes to switch between regimes according to one or more threshold variables. A distinctive feature of T-TFM-cp is its flexibility: all factors may share the same threshold variable, or each factor can be governed by its own thresholding rule. Moreover, unlike the threshold factor models mentioned above, T-TFM-cp incorporates an explicit dynamic specification for the factor processes, thereby enabling prediction, similar in spirit to the dynamic matrix factor model of \cite{YuChen2024}.

In this paper, we extend the TFM-cp model to a nonlinear threshold framework--rather than extending the TFM-Tucker model--for several reasons. 
First, the TFM-cp model \eqref{TFM-cp} is uniquely defined up to sign changes when the signal strengths are ordered, facilitating interpretation. Second, by decomposing the factor processes into uncorrelated univariate time series, TFM-cp enables a more flexible modeling of temporal dynamics. Compared with Tucker-based representations, CP models typically require fewer factors to capture the essential signal structure, achieving a favorable balance between parsimony and explanatory power. Third, the uncorrelated univariate factor structure enables independent modeling of regime-switching dynamics for each factor, potentially with different threshold variables, threshold values, or numbers of regimes. 

By combining CP decomposition with TAR dynamics, the proposed model offers a powerful and flexible framework for analyzing high-dimensional tensor time-series data exhibiting regime-switching behavior in the latent factors. Balancing parsimony, interpretability, and statistical efficiency, it provides a valuable tool for researchers working with complex multiway temporal data in fields such as neuroscience, finance, and signal processing.

The remainder of the paper is organized as follows. Section \ref{S.ModelDescription} presents the proposed model in detail, including the TAR dynamics of the latent factors and the estimation procedures. Section \ref{theory} establishes the theoretical properties of the estimators. Sections \ref{S.Simulations} and \ref{S.RealData} demonstrate the effectiveness of the model through simulation studies and an empirical application, respectively. Section \ref{sec:discussion} concludes with a discussion of future research directions.

\section{Model Setting and Estimation}\label{S.ModelDescription}
    

In this section, we focus on a more specific specification of the T-TFM-cp model introduced in \eqref{TFM-cp} and \eqref{tar0}. In particular, we assume that each latent factor follows a threshold autoregressive (TAR) process of the form
\begin{align} \label{TAR}
    f_{j,t}=\begin{cases}
        \phi^{(1)}_{j,1}f_{j,t-1}+\cdots+\phi^{(1)}_{j,p_j }f_{j,t-p_j}+\varsigma^{(1)}_j \xi_t & \text{if } z_{j,t} \in (-\infty, s_{j,1}],\\
        \phi^{(2)}_{j,1}f_{j,t-1}+\cdots+\phi^{(2)}_{j,p_j }f_{j,t-p_j}+\varsigma^{(2)}_j \xi_t & \text{if } z_{j,t} \in (s_{j,1}, s_{j,2}],\\
        & \vdots \\
        \phi^{(L)}_{j,1}f_{j,t-1}+\cdots+\phi^{(L)}_{j,p_j }f_{j,t-p_j}+\varsigma^{(L)}_j \xi_t & \text{if } z_{j,t} \in (s_{j,L-1}, \infty),
    \end{cases} 
\end{align}
where $L$ is the number of regimes, $s_{j,l}$ are the threshold values, $p_j$ are the autoregressive orders, $\phi_{j}^{(l)}$ are the autoregressive coefficients for regime $l$, and $\xi_t$ are white noise processes with variances 1 for $l = 1, 2, \cdots, L$. In the Self-Exciting TAR (SETAR) case, the exogenous threshold variable $\bz_t$ is assumed to be past values of $f_{jt}$, e.g. $f_{j,t-\tau_j}$, where $\tau_j$ is the delay parameter. In the following we use the notations 
$\bs_j=(s_{j,1},...,s_{j,L-1})^\top \in \R^{L-1}$, $\bphi_j^{(l)}=(\phi_{j,1}^{(l)},...,\phi_{j,p_j}^{(l)})^\top \in \R^{p_j}$, and $\bphi_j=(\bphi_j^{(1)\top},...,\bphi_j^{(L)\top})^\top \in \R^{L p_j}$.


The TAR model is a powerful and elegant nonlinear extension of the linear autoregressive (AR) model, capable of capturing regime switching driven by an observable threshold variable. It has been widely used to model economic and financial data exhibiting asymmetric dynamics and nonlinear patterns. By allowing different autoregressive behaviors depending on the regime of the threshold variable $z_{j,t-d}$, the model can represent phenomena such as asymmetric cycles, shifts in mean levels, and volatility clustering. This flexibility makes TAR models particularly suitable for characterizing distinct dynamics during economic expansions and recessions \citep{Tong90, Chan&Tong86, tsay1989testing}.

We refer to the combination of the TFM-cp model in \eqref{TFM-cp} and the dynamic TAR specification in \eqref{TAR} as the TAR Tensor Factor Model in CP form (T-TFM-cp). To estimate the parameters of the T-TFM-cp model, we adopt a two-step estimation strategy. In the first step, we estimate the TFM-cp model while ignoring the dynamic structure of the latent factors. This yields estimates of both the 
loading vectors $\mathbf{u}_{jk}$ and the factor process $\widehat{f}_{jt}$. The procedure follows \cite{chen2024estimation}, which combines a composite PCA initialization with an iterative simultaneous orthogonalization (ISO) algorithm. Details and theoretical properties can be found in \cite{chen2024estimation}. Denote the estimated loading vectors by $\hat{\bu}_{jk}$ and the estimated factors by $\widehat f_{jt}=\cX_t\times_{k=1}^K \hat \bv_{jk}^{\top}$, where $\hat{\bv}_{jk}$ is the $j$-th column of $\widehat{\bU}_k(\widehat{\bU}_k^\top\widehat{\bU}_k)^{-1}$, with $\widehat{\bU}_k=(\hat{\bu}_{1k},\cdots,\hat{\bu}_{rk})\in \mathbb{R}^{d_k\times r}$. 
Under this definition, $\widehat f_{jt}$ serves as an estimate of $\lambda_j f_{jt}$ in \eqref{TFM-cp}, implicitly absorbing the signal strength $\lambda_j$. This scaling does not affect the results of the simulation study or the real data analysis; it only changes the magnitude of the estimated factors.

Second, once the factor process estimates $\widehat{f}_{jt}$ are obtained, we fit an individual TAR model to each estimated factor process using standard identification and estimation techniques for TAR models \citep{tong1990non,tsay1989testing,tsay2018nonlinear}. In particular, given the threshold variable, the number of regimes, and the AR orders of each regime, the AR coefficients and threshold values are estimated by least squares. Specifically, for each $j=1,\ldots,r$, the estimates $\hat{\bphi}_j$ and $\hat{\bs}_j,j=1,\ldots,r$ minimizes
\begin{equation} \label{LS-TAR}
\sum_{t=p_j+1}^T
\left(\hat{f}_{j,t}-\sum_{\ell}^L
I(z_{j,t}\in(s_{j,\ell-1},s_{j,\ell}))
\sum_{k=1}^{p_j}\phi_{j,k}^{(\ell)}\hat{f}_{j,t-k}\right)^2.
\end{equation}
Although the estimation of the threshold vector $\bs_j$ is strictly a non-convex optimization problem, it is typically not 
computationally intensive since one only needs to check the observed values of the threshold variable; all values between two consecutive observations of the threshold variable are equivalent \citep{chan1993consistency,hansen2000sample}. Sequential update of the least squares (when adding or dropping one observation) makes exhaustive search computationally efficient, though more advanced algorithms are available \citep{li2016nested,li2011least}.  The number of regimes and the AR orders are typically determined by BIC or similar measures, as well as diagnostic model checking procedures involving residual analysis, testing for remaining nonlinearity, and evaluating forecasting performance \citep{Tong90,hansen2000sample,li2011least,gonzalo2002estimation,chan1998limiting}. 

The two-step estimation strategy is standard for models consisting of multiple components, such as the T-TFM-cp model. Similar procedures have been widely used in the literature on dynamic (vector) factor models; see, for example, \cite{stock2016, Jasiak2001, hallin2016, otto2022approximate}.

Threshold variable determination is essential for building a threshold model. When there is no prior knowledge on the threshold variable, a data-driven procedure is needed in order to search for a suitable one. In standard univariate threshold models, a typical candidate pool is the lag variables \citep{Tong&Lim1980,Tong90,chan1993consistency,Tong2010} and identification is commonly performed by comparing a handful of plausible specifications. When multiple time series are involved, as in our setting, it may be desirable to identify one or a small number of common threshold variables, which can substantially improve model interpretability. When one considers a 
large candidate pool consisting of exogenous variables, a trial-and-error approach can be extremely time consuming, complicated more by the multiple comparison problem at the end. In such cases, the reverse approach proposed in \citep{Wu&Chen2007,liu2016regime} provides an effective alternative. It is also worth noting that different factor processes may rely on different types of threshold variables: some may follow self-exciting dynamics using lagged values of the factor itself, while others may require observable exogenous threshold variables.

\section{Theoretical Properties of the Estimators}\label{theory}
    In this section, we present some theoretical properties of the proposed two-stage estimation procedures.
We introduce some notations first. Let $d=\prod_{k=1}^K d_k ,d_{\max}=\max\{d_1,...,d_K\}$. The matrix Frobenius norm is defined as $\|\bA\|_{\rm F} = (\sum_{ij} a_{ij}^2)^{1/2}$.
Define the spectral norm as
$$ \|\bA\|_{2} =  \max_{\|\bx\|_2=1,\|\by\|_2= 1} \|\bx^{\top} \bA \by\|_2.$$ 
Considering that the loading vector $\bu_{jk}$ of TFM-cp can only be identified with a change in sign, we employ
\begin{align*}
\|\widehat \bu_{jk}\widehat \bu_{jk}^\top  - \bu_{jk} \bu_{jk}^\top \|_{2}=\sqrt{1-(\widehat \bu_{jk}^\top \bu_{jk})^2 } = \sup_{\bz\perp \bu_{jk}}|\bz^\top \widehat \bu_{jk}|    
\end{align*}
to quantify the discrepancy between $\widehat \bu_{jk}$ and $\bu_{jk}$.

To establish the asymptotic properties of the proposed procedures, we impose the following assumptions.

\begin{assumption}\label{asmp:error}
The idiosyncratic noise process $\cE_t$ are independent Gaussian tensors, conditioning on the factor process $\{f_{jt}, 1\le j\le r,t\in\mathbb Z\}$. In addition, there exists some constant $\sigma>0$, such that
\begin{equation*}
\E (\bv^{\top} \text{vec}(\cE_t))^2\le \sigma^2 \|\bv\|_2^2, \quad \forall\,\bv\in\RR^d.
\end{equation*}
\end{assumption}

\begin{assumption}\label{asmp:eigenvalue}
Let $\lambda_{1}\ge\lambda_{2}\ge\cdots\ge\lambda_{r}>0$. Suppose $\lambda_1\asymp \lambda_r\asymp \lambda$.
\end{assumption}

\begin{assumption}\label{asmp:mixing}
Assume the factor process $f_{jt}, 1\le j\le r$, is stationary and strong $\alpha$-mixing in $t$, with $\E f_{jt}^2 <\infty$. Let $\bF_t=(f_{1t},...,f_{rt})^\top$. For any $\bv\in\R^{r}$ with $\|\bv\|_2=1$,
\begin{align}\label{cond2}
\max_t\P\left( \left| \bv^\top \bF_{t} \right| \ge x \right) \le c_1 \exp\left( -c_2x^{\gamma_2} \right),
\end{align}
where $c_1,c_2$ are some positive constants and $0<\gamma_2\le 2$. In addition, 
the mixing coefficient satisfies
\begin{align}\label{cond1}
\alpha(m) \le \exp\left( - c_0 m^{\gamma_1} \right)
\end{align}
for some constant $c_0>0$ and $\gamma_1 >0$, where
\begin{align*}
\alpha(m) = \sup_t\Big\{\Big|\P(A\cap B) - \P(A)\P(B)\Big|:
A\in \sigma(f_{js}, 1\le j\le r, s\le t), B\in \sigma(f_{js}, 1\le j\le r, s\ge t+m)\Big\}.
\end{align*}
\end{assumption}

Assumption \ref{asmp:error} closely resembles the noise conditions found in foundational studies such as \cite{bai2002}, \cite{bai2003}, \cite{lam2011}, \cite{LamYao2012}, and other notable contributions in the factor model literature. For ease of exposition, we assume that the noise tensor is independent across time $t$, hence allowing for weak cross-sectional contemporaneous  dependence. Although introducing a weak temporal correlation among noises, as proposed by \cite{bai2002}, is feasible, it significantly complicates the theoretical framework. Additionally, we adopt the normality assumption for technical convenience, although it can be generalized to accommodate exponential-type tail conditions, as seen in \cite{chen2024estimation}.

Assumption \ref{asmp:eigenvalue} is a standard condition in factor models, such as \cite{chen2024estimation} and \cite{han2024cp}. In the case of strong factor models, we have $\lambda= \sqrt{d}$.
Assumption \ref{asmp:mixing} requires the tail probability of $f_{jt}$ to exhibit exponential decay, which is a standard assumption. In particular, when $\gamma_2 = 2$, this implies that $f_{jt}$ follows a sub-Gaussian distribution. The mixing condition is a well-established assumption that encompasses a wide variety of time series models, including causal ARMA processes with continuous innovations, as discussed in \cite{Tong90, tsay2005analysis, fan2008nonlinear, rosenblatt2012markov, tsay2018nonlinear}, among others.

Next, we impose assumptions for the TAR structure of the factor processes, depending on whether the threshold variable $z_{jt}$ is self-exciting (i.e., the past values of $f_{jt}$) or is an exogenous  observable variable.
If $z_{jt}$ is self-exciting with $z_{jt}=f_{j,t-\tau_j}$, then $\bY_{jt} = (f_{j,t-1},\ldots,f_{j,t-(p_j\vee\tau_j)})$ is a Markov chain. Denote its $m$-step transition probability by $\mathsf{P}^m (y,A)$, where $y \in \R^{p_j\vee \tau_j}$ and $A$ is a Borel set.
If $z_{jt}$ is observable, let $\bY_{jt} = (f_{j,t-1},\ldots,f_{j,t-p_j},z_{jt})$,
Denote its $m$-step transition probability by $\mathsf{P}^m (y,A)$, where $y \in \R^{p_j+1}$ and $A$ is a Borel set.

\begin{assumption}\label{asmp:tar-noise}
The innovation of the latent TAR process \eqref{TAR}, $\xi_{jt}$ are i.i.d. with $\E\xi_{jt}=0$, $\E\xi_{jt}^2=1$ and $\E\xi_{jt}^4<\infty$ for each $1\le j\le r$, and $\xi_{jt}$ has a bounded, continuous and positive density on $\R$.
\end{assumption}

\begin{assumption}\label{asmp:tar-trans}
The Markov Chain $\bY_{jt}$ ($1\le j\le r$) admits a unique invariant measure $\Pi(\cdot)$ such that there exist $C_0>0$ and $\rho\in [0, 1)$, for any $y$ and any $m$, $\|\mathsf{P}^m (y,\cdot)- \Pi(\cdot)\|_{\rm TV} \le C_0(1+\|y\|) \rho^m$, where $\|\cdot \|_{\rm TV}$ and $\| \cdot \|$ denote the total variation norm and the Euclidean norm, respectively.
\end{assumption}

\begin{assumption}\label{asmp:tar-iden}
Assume that
\begin{enumerate}
\item[(i)] (when $z_{jt}$ is self-exciting with $z_{jt}=f_{j,t-\tau_j}$): There exist nonrandom vectors $\bw_{j\ell}=(w_{j,\ell, 1},\ldots,w_{j,\ell, p_j})^\top$ with $w_{j,\ell,\tau_j}=s_{j,\ell}$ such that $(\bphi_{j}^{(\ell)} - \bphi_{j}^{(\ell+1)})^\top \bw_{j\ell } \neq 0$ for
$\ell=1,\ldots,L-1$.

\item[(ii)] (when $z_{jt}$ is an observable variable): Assume that $z_{jt}$ has a bounded, continuous and positive density on $\R$, and is Markovian. There exist vector $\bw_{j\ell}=\E [ (f_{j,t-1},\ldots,f_{j,t-p_j})^\top | z_{jt}= s_{j,\ell} ]$ such that $(\bphi_{j}^{(\ell)} - \bphi_{j}^{(\ell+1)})^\top \bw_{j\ell } \neq 0$ for
$\ell=1,\ldots,L-1$.
\end{enumerate}
\end{assumption}

Assumptions \ref{asmp:tar-noise}, \ref{asmp:tar-trans}, \ref{asmp:tar-iden} are commonly used in theoretical analysis of TAR models.
Under Assumption \ref{asmp:tar-trans}, $\bY_{jt}$ is V-uniformly ergodic with $V(\cdot) = K(1 + \|\cdot\|)$, a condition that is stronger than geometric ergodicity. For a detailed explanation of V-uniform ergodicity, see Chapter 16 in \cite{meyn2012markov}. In the specific case where all threshold variables are lags of $f_{jt}$ and errors are homoscedastic across all regimes, Assumption \ref{asmp:tar-noise} combined with the condition $\max_{\ell} \sum_{i=1}^{p_j} |\phi_{j,i}^{(\ell)}|<1$ suffices for Assumption \ref{asmp:tar-trans} to hold. More details can be found in 
\cite{chan1985use} and \cite{chan1989note}.
Assumption \ref{asmp:tar-iden} implies that the autoregressive function is discontinuous at the thresholds $s_{j,\ell}$. 
In Assumption \ref{asmp:tar-iden}(i), if $\tau_j>p_j$, then $w_{j,\ell,\tau_j}$ may not be a component of $\bw_{j\ell}$. In this scenario, Assumption \ref{asmp:tar-iden}(i) is equivalent to the conditions $\bphi_j^{(\ell)}\neq \bphi_j^{(\ell+1)}$ for $1\le \ell \le L-1$, which are both necessary and sufficient for the identification of all thresholds. Assumption \ref{asmp:tar-iden} (ii) is similar to Assumption 3.6 in \cite{zhang2024least}.

\cite{han2024cp} proposed iterative procedure ISO to estimate the factor loading vectors of TFM-cp using auto-covariance tensor. When the number of factors $r$ is fixed, the procedure achieves convergence rate $\| \widehat\bu_{jk}\widehat\bu_{jk}^{\top} -\bu_{jk} \bu_{jk}^{\top} \|_{2} =O_{\P}(\sigma d_{\max}^{1/2}\lambda^{-1}T^{-1/2} +\sigma^2\lambda^{-2}),1\le k\le K$. Using their iterative procedure, we can further establish the theoretical properties of the estimated latent factor process $\widehat f_{jt}$ in \eqref{TFM-cp}.

\begin{proposition}\label{thm:factors}
Suppose Assumptions \ref{asmp:error}, \ref{asmp:eigenvalue}, \ref{asmp:mixing} hold. Assume $r$ is fixed. Let $\widehat f_{jt}=\lambda^{-1} \cX_t\times_{k=1}^K \widehat \bu_{jk}^{\top}$ be the estimated factors using the iterative procedure in \cite{han2024cp}, assuming the initialization condition is satisfied and the signal strength $\lambda$ is known. Then, with probability at least $1-T^{-c}-\sum_{k=1}^K e^{-d_k}$, 
\begin{align}
\left| \widehat f_{jt} - f_{jt} \right| \le C \left( \frac{\sigma \sqrt{d_{\max}} }{\lambda \sqrt{T} } + \frac{\sigma}{\lambda} \right),  \label{thm-factors:eq1}
\end{align}
for all $1\le t \le T$, and
\begin{align}
&\left| \frac{1}{T-h} \sum_{t=h+1}^T \widehat f_{i,t-h} \widehat f_{jt} - \frac{1}{T-h} \sum_{t=h+1}^T f_{i,t-h} f_{jt} \right| \le C \left(  \frac{\sigma \sqrt{d_{\max}} }{\lambda \sqrt{T} }   + \frac{\sigma^2 }{\lambda^2 }  \right),  \label{thm-factors:eq3}
\end{align}
for all $1\le i,j\le r$ and $0\le h\le T/4$, where $c$ is a positive constant.
\end{proposition}

The proposition establishes a non-asymptotic bound 
for the estimated factors $\widehat f_{jt}$. It demonstrates that consistent factor estimation requires the signal-to-noise ratio ($\lambda/\sigma$) to increase to infinity, ensuring sufficient information about the signal at each time point $t$. As expected, a higher signal-to-noise ratio leads to faster convergence, as stronger factors provide more information in the observed data. Additionally, \eqref{thm-factors:eq3} shows that the error rate between the sample (auto-)covariance of the estimated factors and the true sample (auto-)cross-moment is $o_{\P}(T^{-1/2})$ when $\lambda/\sigma\gg \sqrt{d_{\max}}+T^{1/4}$. In the case of a strong factor model setting \citep{bai2002,lam2011}, where $\lambda/\sigma\asymp \sqrt{d}$, this condition is equivalent to $T\ll d^2$, which generally holds. This suggests that under such conditions, using the estimated factor processes for model building and inference of the TAR component is equivalent to using the underlying true factors, without any efficiency loss. The statistical rates in \eqref{thm-factors:eq1} and \eqref{thm-factors:eq3} provide a basis for further modeling of the estimated factors.



    

The following theorems present the convergence rate and asymptotic normality for the least squares estimators \eqref{LS-TAR} in the second-stage TAR modeling of the latent factor process, using $\widehat f_{jt}$ as the true factor process. 
Theorems \ref{thm:sftar} and \ref{thm:obtar} separately examine the cases of self-exciting threshold variables and observable threshold variables.

\begin{theorem}\label{thm:sftar}
Consider $z_{jt}$ in \eqref{TAR} as self-exciting with $z_{jt}=f_{j,t-\tau_j}$, for some $1\le j\le r$. Suppose Assumptions \ref{asmp:error}, \ref{asmp:eigenvalue}, \ref{asmp:mixing}, \ref{asmp:tar-noise}, \ref{asmp:tar-trans}, \ref{asmp:tar-iden}(i) hold. Assume $r$ is fixed. Let $\widehat f_{jt}=\lambda^{-1} \cX_t\times_{k=1}^K \widehat \bv_{jk}^{\top}$ be the estimated factors using the iterative procedure in \cite{han2024cp}, assuming the initialization condition is satisfied and the signal strength $\lambda$ is known.
If $\sigma \sqrt{d_{\max}} /(\lambda \sqrt{T} )   + \sigma^2 /\lambda^2 \to 0$ as $T\to \infty$, then $\widehat\tau_j\to\tau_j$ in probability.
Moreover, 
\begin{align}
\widehat \bs_j - \bs_j &= O_{\P} \left( \frac1T   +   \frac{\sigma \sqrt{d_{\max}} }{\lambda \sqrt{T} }   + \frac{\sigma }{\lambda } \right)  , \label{thm_sftar_eq1}\\
\widehat \bphi_j -\bphi_j &= O_{\P} \left( \frac{1}{\sqrt{T}}   +   \frac{\sigma \sqrt{d_{\max}} }{\lambda \sqrt{T} }   + \frac{\sigma }{\lambda } \right)  .   \label{thm_sftar_eq2}
\end{align}
Furthermore, if the factor strength satisfies $\lambda/\sigma \gg T^{1/2}+ \sqrt{d_{\max}}$, then
\begin{align}
\sqrt{T} \left( \widehat \bphi_j -\bphi_j \right)  \Rightarrow N(0, \bSigma_j),  \label{thm_sftar_eq3}
\end{align}
where $\bSigma_j=\text{diag}\big\{(\varsigma_j^{(1)})^2\Sigma_{j1},...,(\varsigma_j^{(L)})^2\Sigma_{jL}\big\}$ with $\Sigma_{j,\ell}^{-1}=\E [\widetilde \bff_{j,t-1} \widetilde \bff_{j,t-1}^{\top} \ones\{s_{j,\ell-1} < f_{j,t-\tau_j} \le s_{j,\ell} \} ]$, $1\le \ell\le L$, and $\widetilde \bff_{j,t-1} =(f_{j,t-1},...,f_{j,t-p_j})^\top$, $s_{j,0}=-\infty, s_{j,L}=+\infty$, and $\varsigma_j^{(\ell)}$ is defined in \eqref{TAR}.

\end{theorem}

\begin{theorem}\label{thm:obtar}
Consider $z_{jt}$ in \eqref{TAR} as an observable variable, for some $1\le j\le r$. Suppose Assumptions \ref{asmp:error}, \ref{asmp:eigenvalue}, \ref{asmp:mixing}, \ref{asmp:tar-noise}, \ref{asmp:tar-trans}, \ref{asmp:tar-iden}(ii) hold. Assume $r$ is fixed. Let $\widehat f_{jt}=\lambda^{-1} \cX_t\times_{k=1}^K \widehat \bv_{jk}^{\top}$ be the estimated factors using the iterative procedure in \cite{han2024cp}, assuming the initialization condition is satisfied and the signal strength $\lambda$ is known.
Then,
\begin{align}
\widehat \bs_j - \bs_j &= O_{\P} \left( \frac1T  \right)  ,   \label{thm_obtar_eq1}\\
\widehat \bphi_j -\bphi_j &= O_{\P} \left( \frac{1}{\sqrt{T}}   +   \frac{\sigma \sqrt{d_{\max}} }{\lambda \sqrt{T} }   + \frac{\sigma^2 }{\lambda^2 } \right)  .   \label{thm_obtar_eq2}
\end{align}
Furthermore, if the factor strength satisfies $\lambda/\sigma \gg T^{1/4}+ \sqrt{d_{\max}}$, then
\begin{align}
\sqrt{T} \left( \widehat \bphi_j -\bphi_j \right)  \Rightarrow N(0, \bSigma_j),   \label{thm_obtar_eq3}
\end{align}
where $\bSigma_j=\text{diag}\big\{(\varsigma_j^{(1)})^2\Sigma_{j1},...,(\varsigma_j^{(L)})^2\Sigma_{jL}\big\}$ with $\Sigma_{j,\ell}^{-1}=\E [\widetilde \bff_{j,t-1} \widetilde \bff_{j,t-1}^{\top} \ones\{s_{j,\ell-1} < z_{jt} \le s_{j,\ell} \} ]$, $1\le \ell\le L$, and $\widetilde \bff_{j,t-1} =(f_{j,t-1},...,f_{j,t-p_j})^\top$, $s_{j,0}=-\infty, s_{j,L}=+\infty$, and $\varsigma_j^{(\ell)}$ is defined in \eqref{TAR}.

\end{theorem}

In the case of self-exciting threshold variables, the estimation error of the threshold level $\bs_j$ in \eqref{thm_sftar_eq1} comprises two components: the conventional $1/T$ rate from \cite{chan1993consistency,li2012least} and \cite{zhang2024least} and the estimation error of the factor process from the first stage in \eqref{thm-factors:eq1}. This additional error arises because the lagged estimated factor process serves as the threshold variable. Since the estimated factors determine the Markovian regime at each time $t$, we require a non-asymptotic convergence rate for all $\widehat f_{jt}$, as established in Proposition \ref{thm:factors}. The convergence rate provided by Theorem 3 in \cite{han2024cp} is insufficient for this purpose. In contrast, for observable threshold variables, the threshold level $\bs_j$ in \eqref{thm_obtar_eq1} achieves the standard $1/T$ rate without additional factor estimation error.

The presence of factor estimation error in the self-exciting threshold variable case leads to larger AR coefficient estimation errors compared to the observable threshold variable case. For observable thresholds, the error consists of the parametric rate $T^{-1/2}$ and the error from the sample covariance matrix using estimated factors as in \eqref{thm-factors:eq3}. For self-exciting thresholds, an additional term $\sigma/\lambda$ appears, reflecting the signal-to-noise ratio. Consequently, the least squares estimator $\widehat \bphi_j $ satisfies a central limit theorem only when the signal strength is sufficiently high: specifically, $\lambda/\sigma\gg \sqrt{d_{\max}}+ T^{1/4}$ for observable threshold variables and $\lambda/\sigma\gg \sqrt{d_{\max}}+ T^{1/2}$ for self-exciting threshold variables. When the signal is weaker, consistency remains achievable, but the factor loading and idiosyncratic noise $\cE_t$ dominate the estimation error. In such regimes, the factor process error in \eqref{thm-factors:eq3} and/or \eqref{thm-factors:eq1} dominates the parametric rate, preventing the derivation of a tractable asymptotic distribution.

\section{Simulations}\label{S.Simulations}
    
We conducted a simulation study to assess the performance of the proposed T-TFM-cp model in matrix form across a range of settings. The scenarios systematically vary the dimensions, sample sizes, and signal-to-noise ratios (SNR, $\lambda/\sigma$). Specifically, we generate data from weak factor model with $\lambda=1$,
\begin{equation}
\bX_t=\bM_t+\bE_t=\sum_{j=1}^3\bu_{1,j}
\bu_{2,j}'f_{j,t}+\bE_t , \label{simuTFM}
\end{equation}
where, for $j=1,2,3$, the components $\boldsymbol{u}_{1,j}$ and $\boldsymbol{u}_{2,j}$ are generated by drawing i.i.d. Gaussian entries with zero mean and then normalizing the resulting vectors. All elements of $\bE_t$ have variance $\sigma^2$. The latent factors follow the TAR models 

\begin{align}
    f_{1,t}&=\begin{cases}
        0.5f_{1,t-1}+0.2f_{1,t-2}+a_{1,t},& \text{if } f_{1,t-1}<0, \\
        0.7f_{1,t-1}-0.6f_{1,t-2}+a_{1,t},\hspace{3mm} & \text{if } f_{1,t-1}\geq0,
    \end{cases} \nonumber\\ 
        f_{2,t}&=\begin{cases}
        \ \ 0.8f_{2,t-1}+0.1f_{2,t-2}+a_{2,t},& \text{if } f_{2,t-1}<0,\\
        -0.4f_{2,t-1}-0.6f_{2,t-2}+a_{2,t},& \text{if } f_{2,t-1}\geq0,
    \end{cases} \label{simuTAR} \\ 
        f_{3,t}&=\begin{cases}
        \ \ 0.7f_{3,t-1}+a_{3,t},& \text{if } f_{3,t-1}<0,\\
        -0.8f_{3,t-1}+a_{3,t},\hspace{18mm} & \text{if } f_{3,t-1}\geq0,
    \end{cases} \nonumber
\end{align}
where $a_{j,t}$ are independent Gaussian white noise with zero mean and unit variance ($\varsigma_j^2 = 1$), for $j = 1, 2, 3$. All results are based on 100 simulation replications.

\begin{figure}[ht]
\centering
\includegraphics[width=0.8\linewidth]{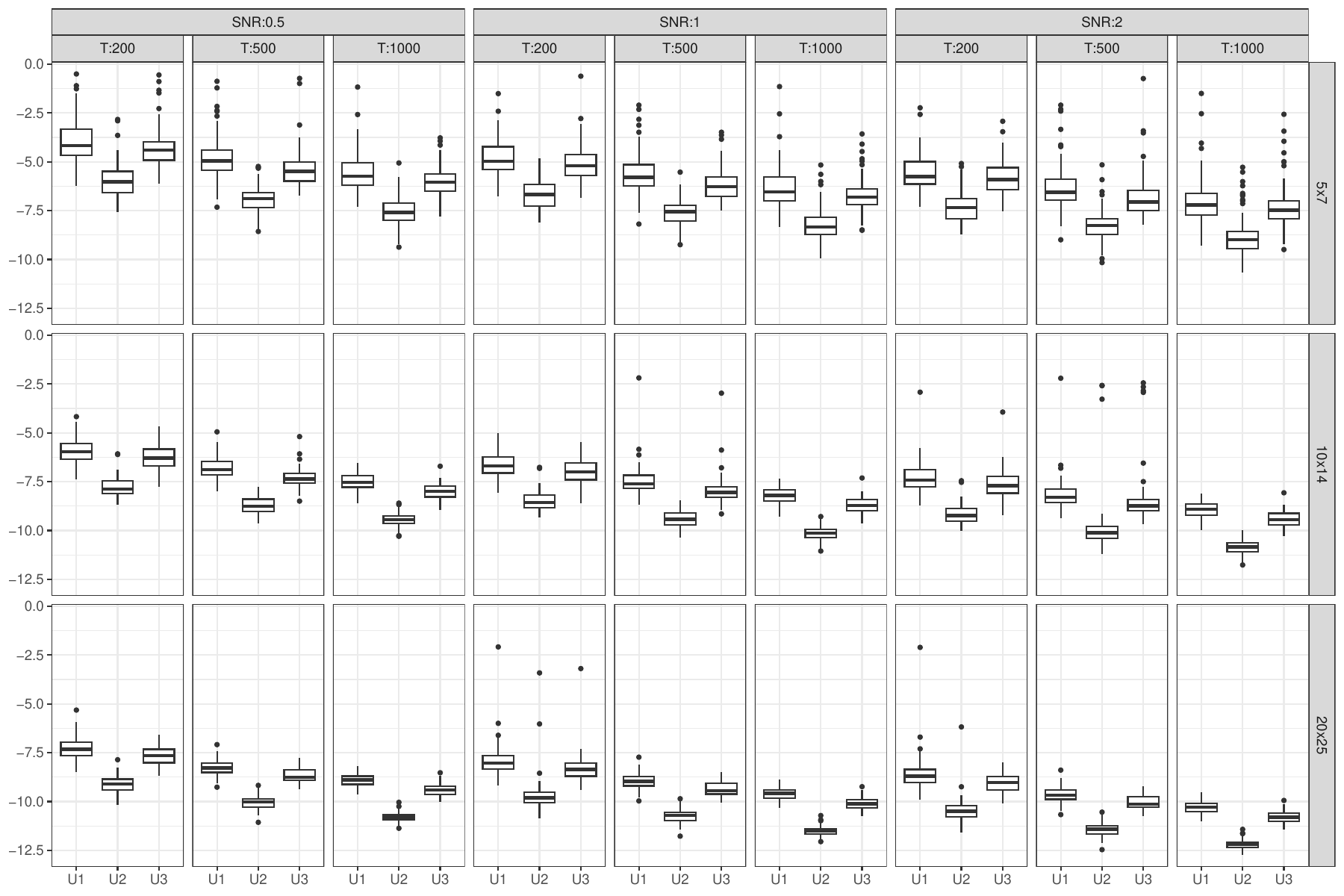}
\caption{Boxplot of log of MSE for estimating the loading vectors ($j=1,2,3$), for combinations of matrix dimension ($5\times 7$, $10\times 14$, $20\times 25$, SNR level ($0.5,1,2$) and time series length ($200,500,1000$).}
\label{fig:MSEuvj}
\end{figure}

Given that the TFM-cp factors are identifiable only up to sign changes and permutations, we first aligned the estimated factors with the true simulated factors using optimal sign matching and permutation selection. We then evaluate the estimation accuracy of the loading vectors using the Mean Squared Error (MSE):
\begin{align}
\text{MSE}_j = \frac{|\widehat{\boldsymbol{u}}_{1,j} - \boldsymbol{u}_{1,j}|_2^2}{d_1} + \frac{|\widehat{\boldsymbol{u}}_{2,j} - \boldsymbol{u}_{2,j}|_2^2}{d_2}, \quad j = 1, 2, 3. \nonumber
\end{align}
We use the estimation procedure of \cite{han2024cp} to estimate the TFM-cp model, using lag 1 ($h=1$) auto-moments in the estimation. Figure \ref{fig:MSEuvj} presents boxplots of the logarithm of $\text{MSE}_j$ for the three loading vectors, under three choices of matrix dimensions, three SNR levels, and three time-series lengths. As expected, the estimation error decreases as the sample size, dimensionality, and SNR increase.

\begin{figure}[ht]
\centering
\includegraphics[width=0.8\linewidth]{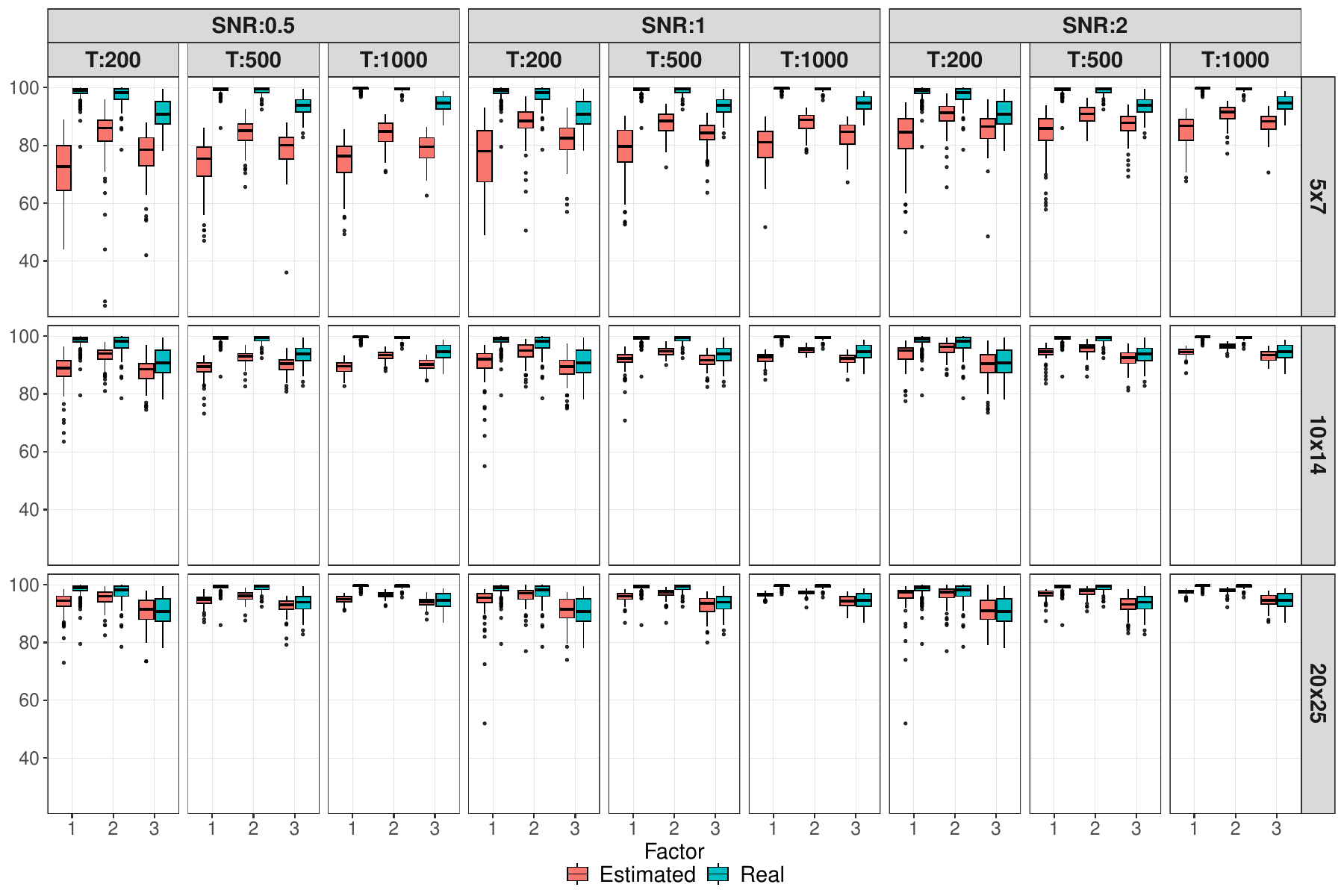}
\caption{Proportion of correctly identified threshold regimes using the estimated factor process (marked as `Estimated') versus the true factor process (marked as `Real').}
\label{fig:MSE_TARregion}
\end{figure}

Next we study the impact of using the estimated factor process as well as its lag variable as the threshold variable, compared to using the underlying true
factor process and its lag threshold variable. To isolate the impact, we fix the true TFM-cp rank $r=3$, use the true AR order each factor as in \eqref{simuTAR}, and set the threshold delay to its true value ($\tau_j=1$). Treating the estimated $\hat{f}_{j,t}$ ($j=1,2,3$) as observed time series, we estimate the threshold model (the AR parameters and the threshold value) under the true
configuration and the estimated threshold variable $\hat{f}_{j,t-1}$.
Given the estimated threshold variable $\hat{f}_{j,t-1}$ and the estimated threshold $\hat{s}_{1,j}$, each observation $\hat{f}_{j,t}$ is classified into one of the two regimes (lower vs. upper). We then compute the proportion of observations for which the inferred regime matches the underlying true regime (i.e. $f_{i,t-1}<0$ for the lower regime and $f_{i,t-1}\geq 0$ for the upper regime). 
For comparison, we repeat the procedure using the ``true'' factor process $f_{i,t}$ and the ``true'' threshold variable $f_{i,t-1}$ to estimate the AR coefficients and the threshold $\tilde{s}_{1,j}$. Using $f_{j,t-1}$ and $\tilde{s}_{1,j}$, we compute the corresponding correct-classification proportion. Figure \ref{fig:MSE_TARregion} shows the comparison of these two proportions under the same model settings as Figure~\ref{fig:MSEuvj}. As expected, the results based on the true factor process do not vary across different dimensions or SNR levels, since they are independent of the first-stage TFM-cp estimation. In contrast, the results based on the estimated factor processes approach the performance of the true-factor case as the dimension or SNR increases, reflecting the improved accuracy of the first-stage estimation.

\begin{figure}[ht]
\centering
\includegraphics[width=0.6\linewidth]{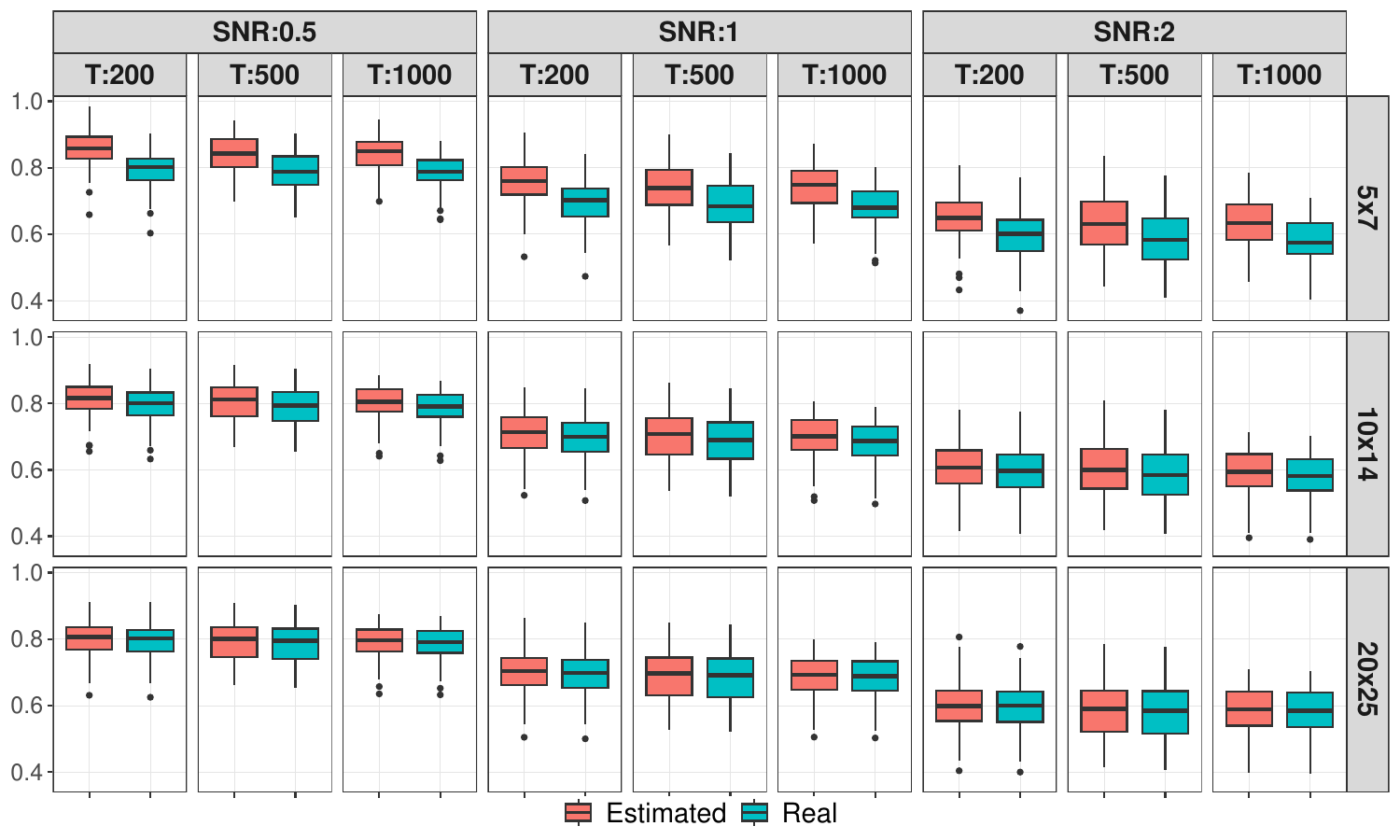}
\caption{Boxplot of one-step ahead prediction error MSE for the prediction of the observations $\mathbf{X}_t$ for $t=T+1,\ldots, T+200$. The boxes marked as `Estimated' is obtained using the estimated loading vectors and 
estimated threshold model parameters obtained at $T$. The boxes
marked as `Real' is obtained using the underlying true loading
vectors and true factor processes.}

\label{fig:PredErrorY}
\end{figure}

Prediction of $\bX_{t+1}$ given $\bX_{1},\ldots, \bX_{t}$ is carried out by first forecasting the factor process. To assess the prediction performance, we generated 200 additional observations for each simulation replicate beyond the original sample size $T$.
Without updating the estimated model parameters, rolling one-step ahead forecasts are produced, and prediction accuracy is assessed using the sum of squared prediction errors
\[
\sum_{t=T}^{T+199} ||\widehat{\bX}_t(1)-\bX_{t+1}||^2_F.
\]
Here, for $t\geq T$, $\widehat{\bX}_t(1)=\sum_{j=1}^3\hat{\bu}_{1,j} \hat{\bu}_{2,j}'\hat{f}_{j,t}(1)$, where $\hat{f}_{j,t}(1)$ denotes the one-step-ahead forecast from the estimated TAR model. These forecasts are based on the TFM-cp factor estimates $\hat{f}_{1,t}, \hat{f}_{2,t}, \hat{f}_{3,t}$, obtained by projecting $\bX_{t}$ onto the space spanned by the estimated loading vectors $\hat{\bu}_{1,j}, \hat{\bu}_{2,j}$, $j=1,2,3$, which were estimated at $T$. For comparison, prediction errors were also computed using the underlying true factor processes $f_{j,t}$ to build the TAR model and forecast $f_{j,t+1}$, which leads to predictions of $\bX_{t+1}$. 
Figures \ref{fig:PredErrorY} presents the box plots of the prediction errors, comparing the prediction using estimated loading vectors and the estimated factor process (marked as ``Estimated"), and that using the underlying true loading vectors and the underlying true factor processes (marked as ``Real"). 
The results exhibit a pattern similar to that in Figure~\ref{fig:MSE_TARregion}: when the dimension of $\bX_t$ is large, the prediction performance based on estimated quantities approaches that obtained using the true underlying factors.

\begin{figure}[ht]
\centering
\includegraphics[width=0.6\linewidth]{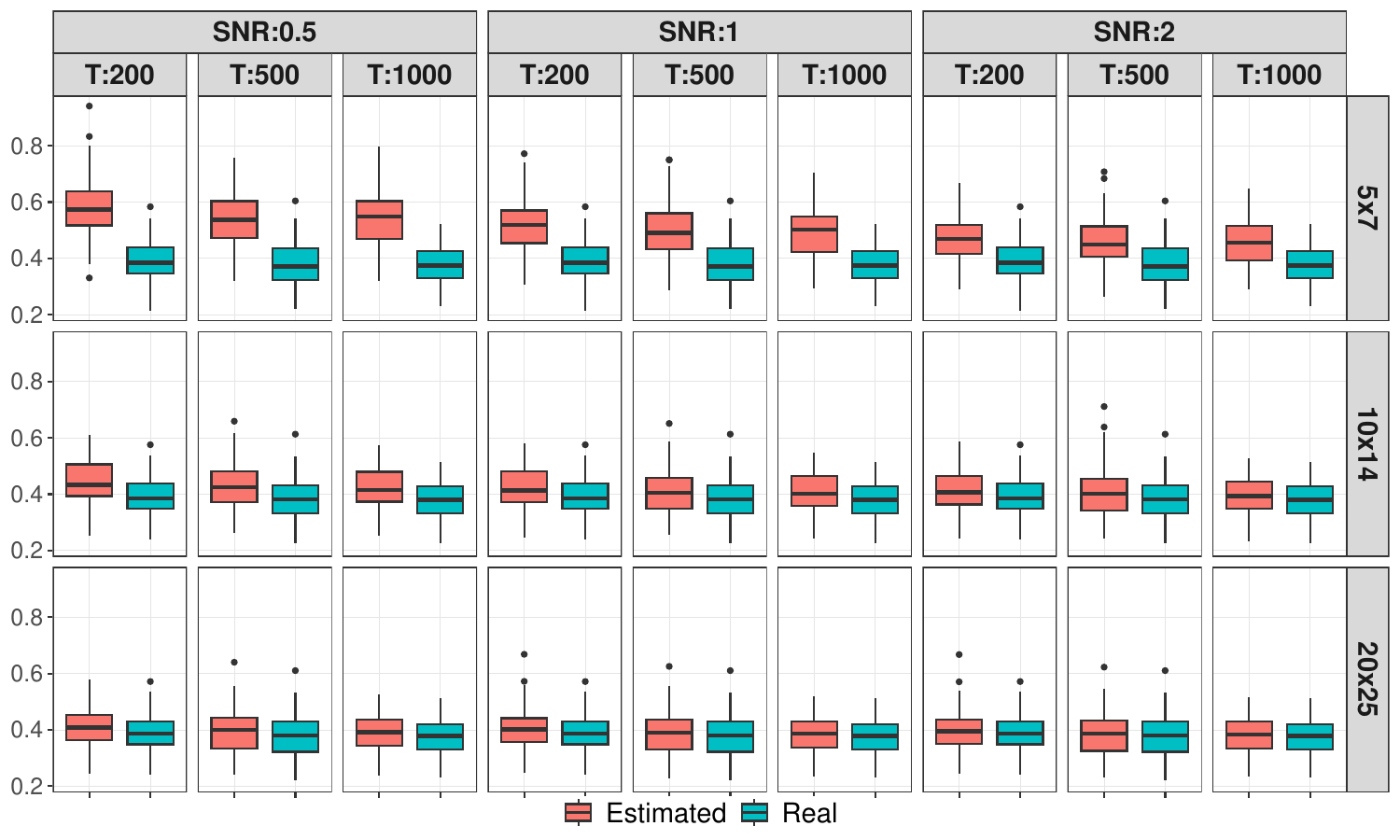}
\caption{Boxplot of one-step ahead prediction error MSE for the prediction of the signal $\mathbf{M}_t$ for $t=T+1,\ldots, T+200$. The boxes marked as `Estimated' is obtained using the estimated loading vectors and 
estimated threshold model parameters obtained at $T$. The boxes
makred as `Real' is obtained using the underlying true loading
vectors and true factor processes.}

\label{fig:predErrorX}
\end{figure}

To eliminate the impact of the unpredictable noise in the observed $\bX_{t+1}$, Figure~\ref{fig:predErrorX} shows the performance of 
\[
\sum_{t=T}^{T+199} ||\hat{\bX}_t(1)-\bM_{t+1}||^2_F,
\]
where $\bM_t$ is the signal part in \eqref{simuTFM}. The difference between the two approaches becomes slightly more pronounced, but the overall pattern remains consistent.

These simulation results demonstrate the effectiveness of our proposed approach in accurately estimating the loading vectors, identifying regime switching in latent factors, and predicting future observations in tensor time series data. Across a wide range of model settings, including varying dimensionality, sample size, and signal-to-noise ratios, the method consistently delivers reliable estimation and forecasting performance.
 \section{Real data example: Multi-country economic indeces}\label{S.RealData}

In this section, we demonstrate the proposed T-TFM-cp model by analyzing the matrix time series containing several economic indices from several countries.


\subsection{Data}

We use the dataset from \cite{ChenChenTsay2019}, which contains quarterly observations from 1990Q4 to 2016Q4 from the 14 countries including: United States of America (USA), Canada (CAN), New Zealand (NZL), Australia (AUS), Norway (NOR), Ireland (IRL), Denmark (DNK), United Kingdom (GBR), Finland (FIN), Sweden (SWE), France (FRA), Netherlands (NLD), Austria (AUT), Germany (DEU). Our analysis focuses on ten macroeconomic indicators: CPGDFD.d2lnsa, CPGREN.d2lnsa, CPALTT01.d2lnsa, IRLT.dlv, IR3TIB.dlv, PRINTO01.dln, PRMNTO01.dln, LORSGPOR.dln, XTEXVA01.GP, and XTIMVA01.GP.

Table \ref{tab:data-transformations} summarizes each series, including its short name, its mnemonic (the series label used in the OECD database), the transformation applied to the series, and a brief description of the data. All data are obtained from the OECD Database. In the transformation column, $\Delta$ denotes the first difference, and $\Delta \ln$ denotes the first difference of the logarithm. GP denotes the growth rate measure from the last period.

\begin{table}[ht]
    \centering
    {\small
    \begin{tabular}{llll}
        \toprule
        \textbf{Short name} & \textbf{Mnemonic} & \textbf{Tran} & \textbf{Description} \\
        \midrule
        CPI: Food & CPGDFD & $\Delta^2 \ln$ & Consumer Price Index: Food, seasonally adjusted \\
        CPI: Ener & CPGREN & $\Delta^2 \ln$ & Consumer Price Index: Energy, seasonally adjusted \\
        CPI: Tot & CPALTT01 & $\Delta^2 \ln$ & Consumer Price Index: Total, seasonally adjusted \\
        IR: Long & IRLT & $\Delta$ & Interest Rates: Long-term government bond yields \\
        IR: 3-Mon & IR3TIB & $\Delta$ & Interest Rates: 3-month Interbank rates and yields \\
        P: TIEC   & PRINTO01 & $\Delta \ln$ & Production: Total industry excluding construction \\
        P: TM & PRMNTO01 & $\Delta \ln$ & Production: Total Manufacturing \\
        GDP     & LQRSGPOR & $\Delta \ln$ & GDP: Original (Index 2010 = 1.00, seasonally adjusted)\\
        IT: Ex & XTEXVA01 & $\Delta \ln$ & International Trade: Total Export Value (Goods) \\
        IT: IM  & XTIMVA01 & $\Delta \ln$ & International Trade: Total import value (goods) \\
        \bottomrule
    \end{tabular}
    }
    \caption{Data transformations and variable definitions}
    \label{tab:data-transformations}
\end{table}

\begin{figure}[ht]
\centering
\includegraphics[width=0.9\linewidth]{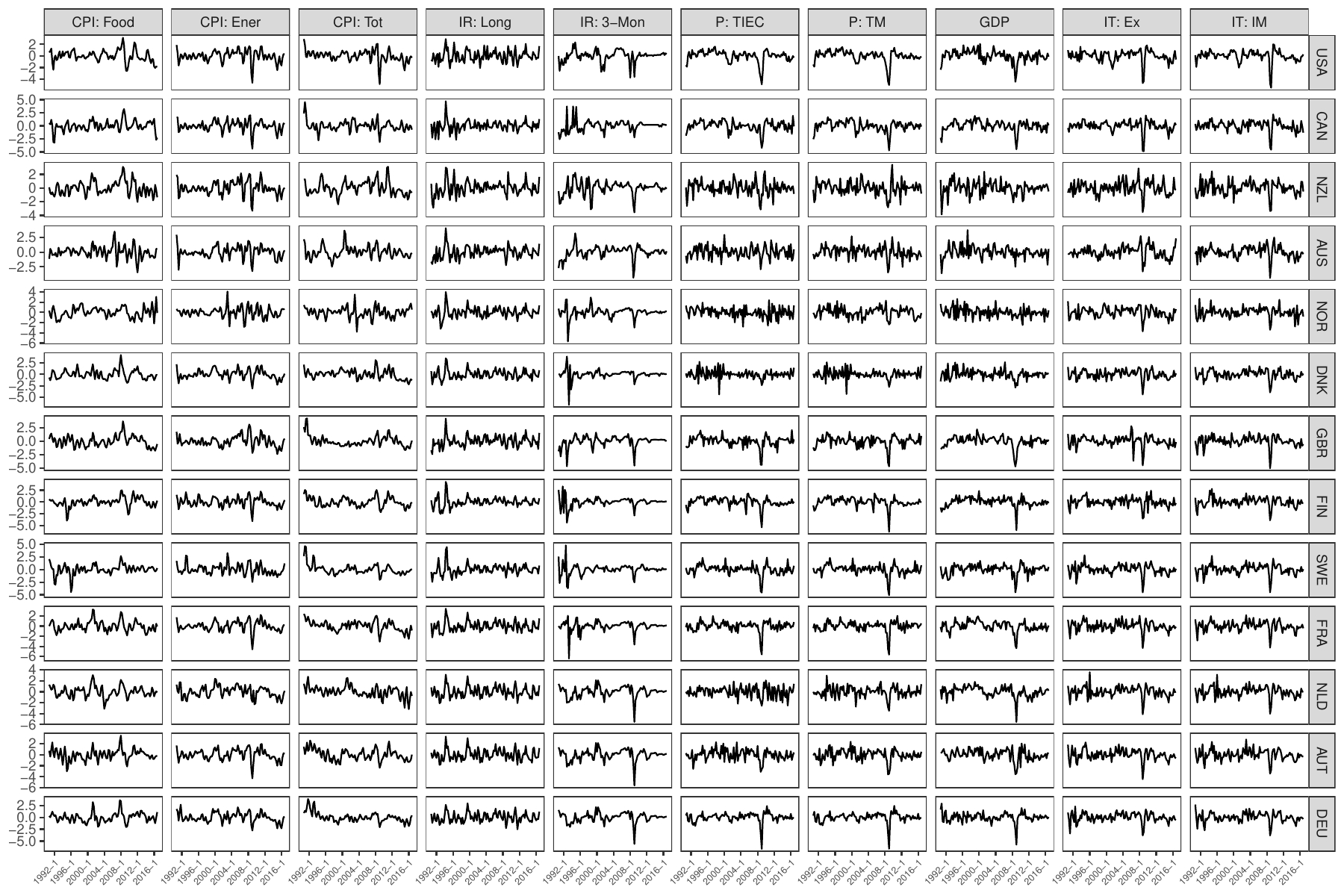}
\caption{Scaled time series, mean zero and $s^2=1$.}
\label{Fig:SeriesYt}
\end{figure}

Figure \ref{Fig:SeriesYt} shows the standardized time series data; the rows and the indices by the columns represent countries.

\subsection{Exploratory Analysis}


As an exploratory analysis, univariate ARMA model is fitted to each individual time series, using
corrected Akaike Information Criterion (AICc) approach \citep{hurvich1989regression,hyndman2008automatic} for model determination. 

\begin{figure}[ht]
\centering
\includegraphics[width=0.7\linewidth]{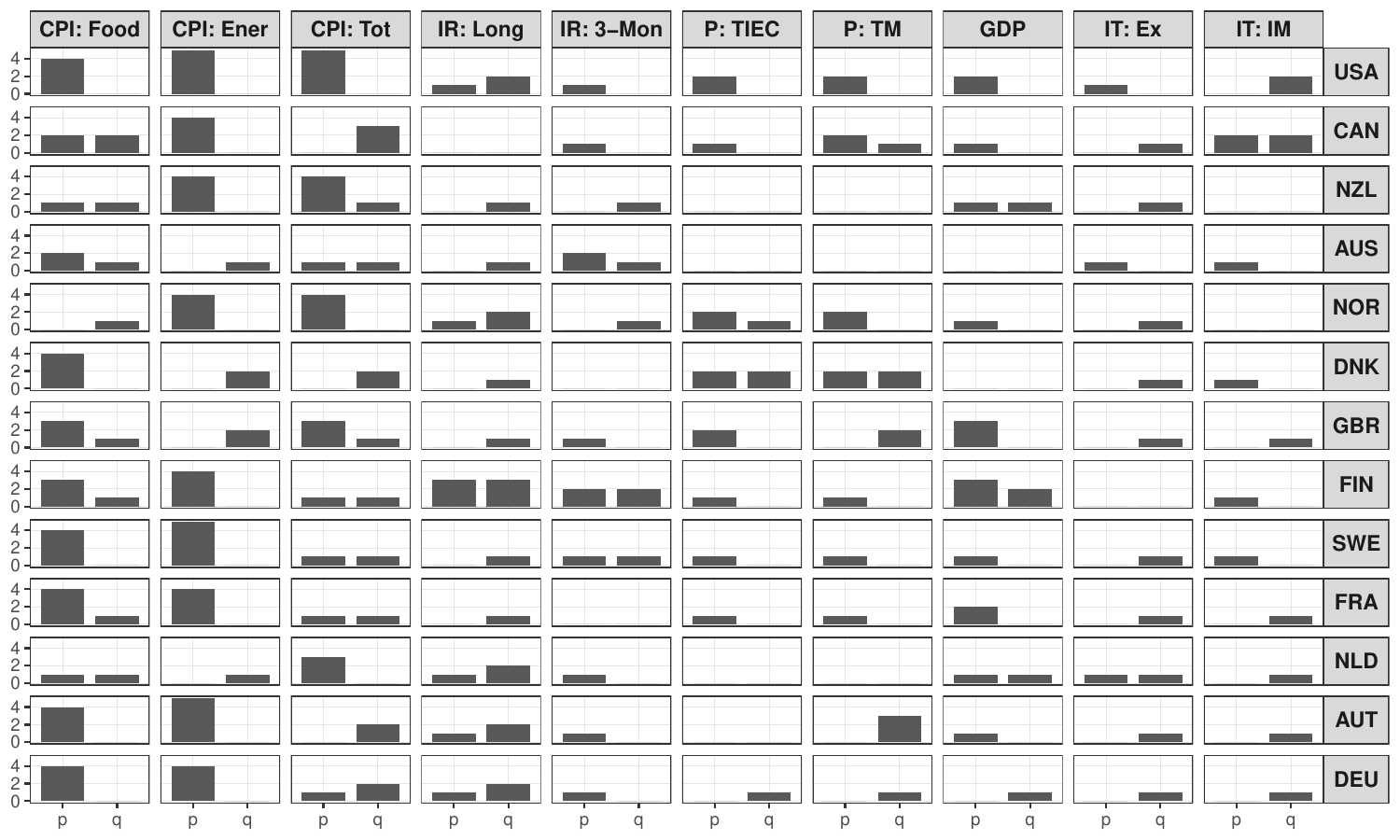}
\caption{Order of $ARMA(p,q)$ model for each individual time series, selected by corrected Akaike Information Criterion (AICc)}
\label{Fig:ARMApq}
\end{figure}

Among the 130 selected univariate ARMA models, 28 are MA(1), 23 are AR(1), 16 are AR of order 4 or 5, 15 are white noise, 10 are ARMA(1,1), and some other less frequently used models. 
Figure \ref{Fig:ARMApq} displays the orders of the ARMA models for each time series selected by AICc.
Notably, the first three indices (CPI related) show a different pattern from the remaining seven indices. 

We stack each observed matrix into a 130 dimensional vector, and estimate a vector factor model (VFM) with two factors using the estimation procedure of 
\cite{LamYao2012}, with $h=1$. 
The number of factors is determined
using the eigen-ratio criterion in
\cite{LamYao2012}.
A TFM-cp model is estimated with two factors, again using the estimation procedure of \cite{han2024cp} with lag 1 ($h=1$) auto-moments. Two factors are used so to be comparable with the VFM model. Note that the loading matrix in the VFM model uses $130\times 2-2$ parameters, while that in TFM-cp uses $(13-1+10-1)\times 2$ parameters for the two (standardized) rank-one $13\times 10$ matrices.  

\begin{figure}[ht]
\centering
\includegraphics[width=0.4\linewidth]{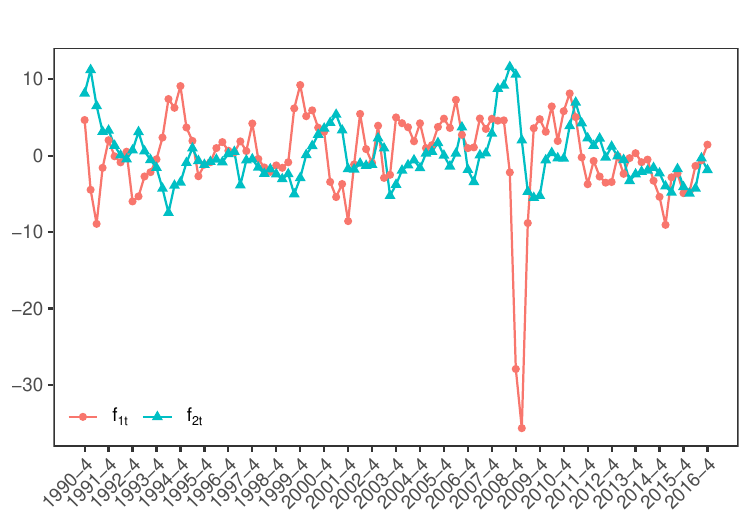}
\includegraphics[width=0.4\linewidth]{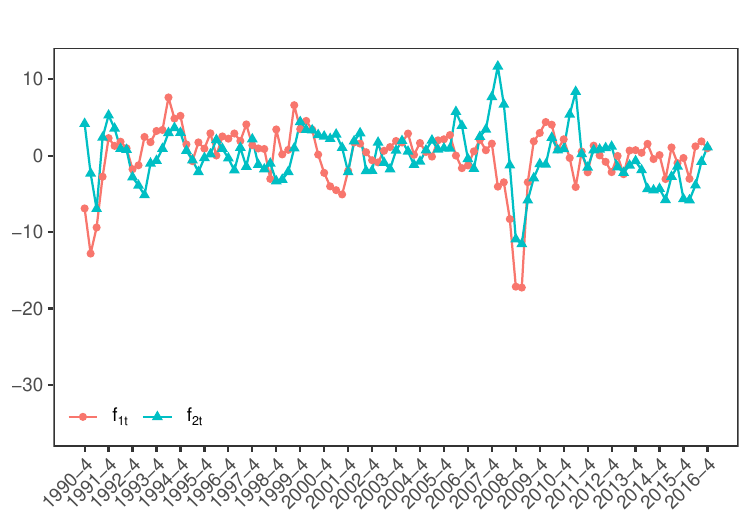}
\caption{Estimated factors using VFM (left) and
TFM-cp (right)}
\label{fig:FactorsFullData}
\end{figure}

Figure \ref{fig:FactorsFullData} shows the estimated factors of VFM (left) and TFM-cp (right).
The two figures are relatively similar, except for the financial crisis period around the beginning of 2009. During the period, VFM tries to capture the extreme event with one factor, while TFM-cp uses both factors with less magnitude.   

\begin{figure}[ht]
\centering
\includegraphics[width=0.45\linewidth]{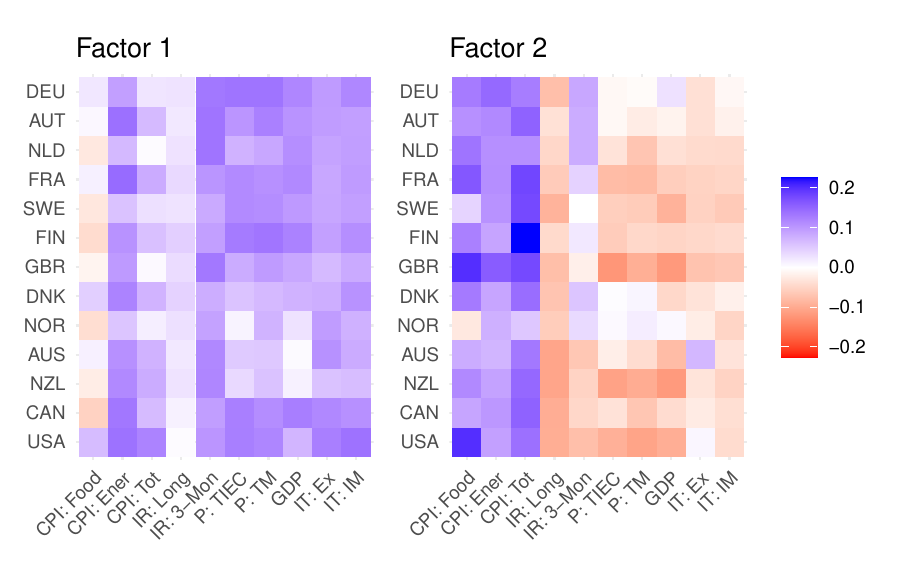}
\includegraphics[width=0.45\linewidth]{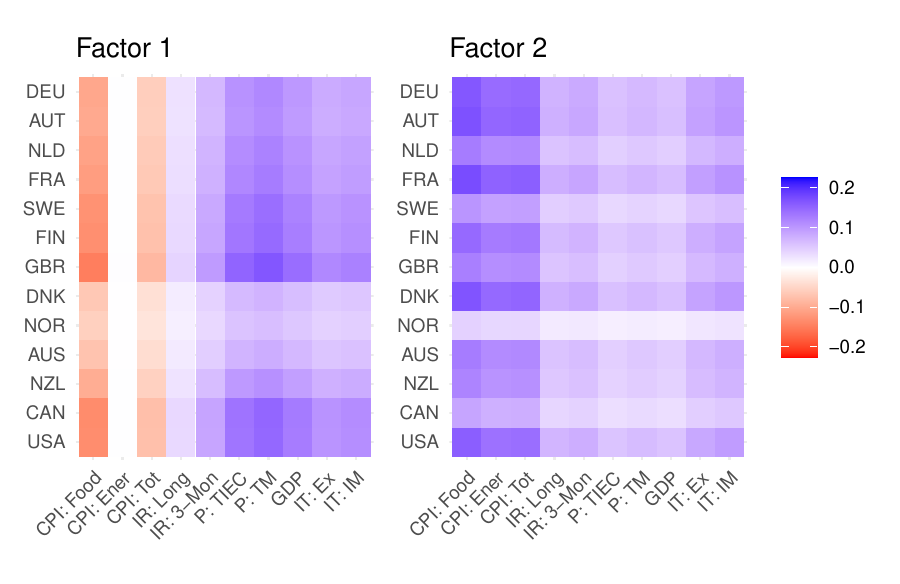}
\caption{Estimated loading matrices. Left: VFM. Right: TFM-cp.}
\label{fig:LoaddingsFullData}
\end{figure}

To further compare the estimated VFM and TFM-cp, 
each column of the estimated loading matrix of the VFM model is rearranged into a matrix according to the column and row classifications of the corresponding observed element in the stacked vector.  
Figure \ref{fig:LoaddingsFullData} shows the rearranged matrix of the VFM loading vectors (left) and the estimated rank-one loading matrices of TFM-cp model (right). 
There are some similarity in both models. For example, the first three indices (CPI related) are loaded heavily on the second factor in both models. Index 4 (long interest rate) is very weakly related to Factor 1, but (negatively) loads on the second Factor in VFM, but not for TFM-cp.  
Most of the indeces have negative loadings on Factor 2 for VFM, 
while the CPI related indices have negative loading on Factor 1 for 
TFM-cp. 
The TFM-cp model show more country-wide differences in the loading. For example, in TFM-cp, DNK, NOR and AUS are very weakly loaded on Factor 1 in TFM-cp, and NOR is weakly loaded on Factor 2.

\begin{figure}[ht]
\centering
\includegraphics[width=0.8\linewidth]{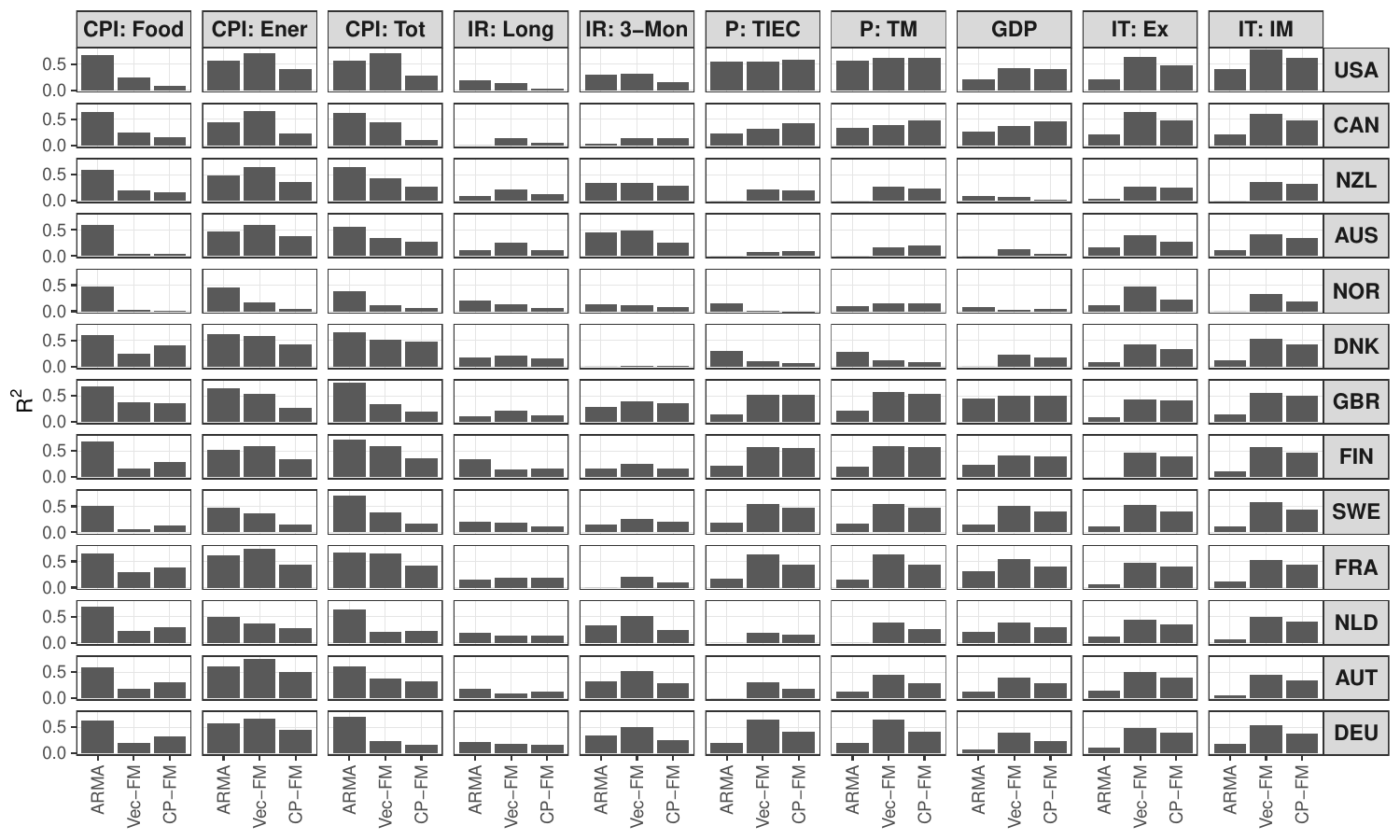}
\caption{In-sample $R^2$ for each time series under the AICc selected ARMA model, VFM and TFM-cp models. 
}
\label{fig:R2FullData}
\end{figure}

For each model, we compute the in-sample fitted values and obtain residuals for each time series. Figure~\ref{fig:R2FullData} reports the in-sample coefficient of determination, $R^2=1-\sum(y_t-\hat{y}_t)^2/\sum y^2_t$ for each series under the best ARMA model selected by AICc shown in Figure~\ref{Fig:ARMApq}, as well as the VFM and TFM-cp specifications. The overall in-sample $R^2$ values for the ARMA, VFM, and TFM-cp models are 0.2993, 0.3839, and 0.2937, respectively, using 258, $257 = 130\times 2 - 3$, and $42 = 2{(10-1)+(13-1)}$ parameters.

Figure~\ref{fig:R2FullData} indicates that the ARMA models provide markedly better fits for the first three CPI related series—albeit with relatively high ARMA orders (see Figure~\ref{Fig:ARMApq})—whereas the factor models yield better fits for indices 5–10. None of the models captures the dynamics of the fourth index (IR: Long) satisfactorily.

More specifically, for the first three series, the overall $R^2$ values are 0.5941 (ARMA), 0.3331 (VFM), and 0.3437 (TFM-cp); for the remaining indices, the corresponding values are 0.1633, 0.3470, and 0.3521. It is worth noting that VFM employs substantially more parameters than TFM-cp.

\subsection{Analysis using T-TFM-cp}

The VFM and TFM-cp do not specify any 
temporal structure of the factors. In the following we present results of fitting ARMA and TAR models for the 
estimated factors under each setting. The results show that there is indeed a threshold phenomenon among the factors. 

Table~\ref{tbl:models65} reports the models estimated using the first 65 observations; the remaining 40 observations are reserved for evaluating out-of-sample predictive performance. Based on preliminary exploration, we use the log growth of U.S. GDP, $z_t= ln(US.GDP_t) - ln(US.GDP_{t-1})$
as the threshold variable for all factors.
U.S. GDP growth is often used as the threshold variable in economic studies as it is a good representative for the status of the economy \citep[e.g.,][]{enders2007threshold,osinska2020modeling,hansen2011threshold,tiao1994some}.
Over the estimation period, the resulting threshold variable has a mean of 0.0130 
and a standard deviation of 0.0050 

The T-TFM-cp estimation results reveal markedly different dynamics across regimes for both factors. Factor 1 exhibits a clear threshold effect at 0.0113 (with delay of $d=4$ quarters), splitting the sample into a relatively small regime with 17 observations and a large regime with 44 observations. The upper regime (Regime 1) shows pronounced dynamics, with a strong $AR(1)$ effect ($0.90$) and a significant negative AR(2) term ($-0.43$), indicating substantial oscillatory behavior. In contrast, Regime 2 is well-described by a simple AR(1) structure with a more moderate coefficient ($0.65$), suggesting a smoother dynamics. The threshold variable used for factor 2 is the same as factor 1, with a smaller threshold value, resulting that the upper regime has more observations. Note that the overall  residual variances and AIC values of both TAR model for factor 1 and factor 2 are much smaller that of fitting the linear ARMA models to the (same) factors under TFM-cp. The overall variances and AIC values under T-TFM-cp is also smaller than that under 
VFM using both ARMA and TAR model for the factor processes.

The estimated models using the full sample size is shown in Table~\ref{tbl:models105} in Appendix Appendix B. 
The T-TFM-cp model under full sample size is very similar to that of using the first 65 observations. Even though the threshold values are slightly different, the proportions of observations in each regime are similar, showing the relative stability of the estimated models. 
%

\begin{figure}[ht]
\centering
\includegraphics[width=0.7\linewidth]{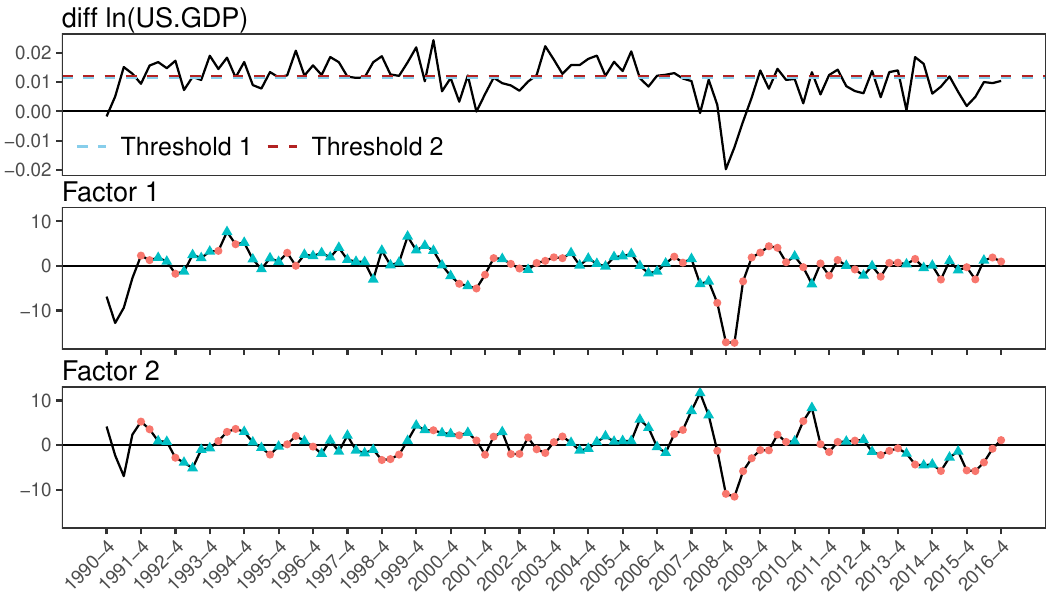}
\caption{The threshold variable and the estimated factors with regime labels of the T-TFM-cp model. 
}
\label{fig:LabelsTAR-cp-fm}
\end{figure}

Figure~\ref{fig:LabelsTAR-cp-fm} illustrates the regimes identified by the T–TFM-cp model. Red dots indicate Regime~1, and blue triangles indicate Regime~2. The top panel shows the threshold variable for reference, while the middle and bottom panels display Factors 1 and 2, respectively. The two factors exhibit slightly different thresholds, with values of 0.0113 for Factor 1 and 0.0121 for Factor 2, highlighting subtle differences in their regime separation.


To assess the performance of each model
listed in Table~\ref{tbl:models65}, 
out-sample rolling 1-step prediction errors are obtained for the last 40 observations in the data set, from first quarter of 2007 to forth quarter of 2016. 
In the rolling prediction, the structure of the models (e.g. number of factors, the estimated loading vectors, 
the ARMA orders or the
Threshold AR orders, and threshold 
variable, the number of regimes and the threshold values) 
are fixed as listed in
Table~\ref{tbl:models65}. Other model parameters, including the factors in the prediction period, the
parameters in ARMA or the TAR, are updated using observations 1 to $t$ for the prediction of $f_{j,t+1}$ and $\bX_{t+1}$. 

\begin{table}[ht]
\centering
{\small
\begin{tabular}{|p{1.6cm}|c|l|l|c|c|c|}
\hline
\textbf{Model} & \textbf{Series} & \textbf{Regime} & \textbf{Coefficients} & \textbf{Threshold} & $\sigma^2$ & \textbf{AIC} \\ \hline
\multirow{2}{*}[-0.5em]{\parbox{1.8cm}{\centering \textbf{TFM-cp ARMA}}}
 & factor1 & -- &
 \begin{tabular}[c]{@{}l@{}}ar1: 0.72 (0.09)\end{tabular}
 & -- & 5.77 & 302.13 \\ \cline{2-7}
 & factor2 & -- &
 \begin{tabular}[c]{@{}l@{}}ma1: 0.68 (0.15)\\ma2: 0.16 (0.15)\end{tabular}
 & -- & 4.74 & 290.06 \\ \hline
\multirow{2}{*}[-0.5em]{\parbox{1.6cm}{\centering \textbf{VFM ARMA}}}
 & factor1 & -- &
 \begin{tabular}[c]{@{}l@{}}ma1: 0.71 (0.12)\\ma2: 0.18 (0.11)\end{tabular}
 & -- & 10.58 & 342.25 \\ \cline{2-7}
 & factor2 & -- &
 \begin{tabular}[c]{@{}l@{}}ar1: 0.56 (0.16)\\ma1: 0.42 (0.20)\end{tabular}
 & -- & 4.37 & 285.23 \\ \hline
\multirow[c]{4}{*}[-2em]{\parbox{1.6cm}{\centering \textbf{TFM-cp TAR}}}
 & \multirow{2}{*}{factor1} & \textbf{Regime 1} (17 obs) &
 \begin{tabular}[c]{@{}l@{}}ar1: 0.90 (0.17)\\ar2: -0.43 (0.16)\end{tabular}
 & \multirow{2}{*}{\parbox{1.5cm}{\centering 0.0113 ($d=4$)}} 
 & \multirow{2}{*}{4.18}
 & \multirow{2}{*}{88.07} \\ \cline{3-4}
 &  & \textbf{Regime 2} (44 obs) &
 \begin{tabular}[c]{@{}l@{}}ar1: 0.65 (0.13)\end{tabular}
 &  &  &  \\ 
 \cline{2-7}
 & \multirow{2}{*}{factor2} & \textbf{Regime 1} (25 obs) &
 \begin{tabular}[c]{@{}l@{}}ar1: 1.1 (0.21)\\ar2: -0.67 (0.19)\\ar3: 0.50 (0.20)\\ar4: -0.58 (0.18)\end{tabular}
 & \multirow{2}{*}{\parbox{1.5cm}{\centering 0.0121 ($d=4$)}}
 & \multirow{2}{*}{3.25}
 & \multirow{2}{*}{76.49} \\ \cline{3-4}
 &  & \textbf{Regime 2} (36 obs) &
 \begin{tabular}[c]{@{}l@{}}ar1: 0.57 (0.12)\end{tabular}
 &  &  &  \\ \hline
\multirow{4}{*}[-0.5em]{\parbox{1.6cm}{\centering \textbf{VFM TAR}}}
 & \multirow{2}{*}{factor1} & \textbf{Regime 1} (38 obs) &
 \begin{tabular}[c]{@{}l@{}}ar1: 0.61 (0.15)\\ar2: -0.30 (0.18)\\ar3: 0.33 (0.17)\\ar4: -0.56 (0.16)\end{tabular}
 & \multirow{2}{*}{\parbox{1.5cm}{\centering 0.0147 ($d=4$)}}
 & \multirow{2}{*}{7.62}
 & \multirow{2}{*}{128.21} \\ \cline{3-4}
 &  & \textbf{Regime 2} (23 obs) &
 \begin{tabular}[c]{@{}l@{}}ar1: 0.88 (0.14)\end{tabular}
 &  &  &  \\ \cline{2-7}
 & \multirow{2}{*}{factor2} & \textbf{Regime 1} (53 obs) &
 \begin{tabular}[c]{@{}l@{}}ar1: 0.63 (0.13)\\ar2: -0.10 (0.13)\end{tabular}
 & \multirow{2}{*}{\parbox{1.5cm}{\centering 0.0179 ($d=2$)}}
 & \multirow{2}{*}{3.41}
 & \multirow{2}{*}{80.15} \\ \cline{3-4}
 &  & \textbf{Regime 2} (10 obs) &
 \begin{tabular}[c]{@{}l@{}}ar1: 1.09 (0.22)\end{tabular}
 &  &  &  \\ \hline
\end{tabular}
}
\caption{Estimated parameters of ARMA and TAR models fitted to the factor process $\widehat{f}_{1t}$, based on the first 65 observations (indices 1–10). 
}
\label{tbl:models65}
\end{table}

{Table \ref{tbl:predError_Split}} 
shows the prediction MSEs over the prediction period 
under five 
different models (individual ARMA, ARMA-VFM, ARMA-TFM-cp, T-VFM and T-TFM-cp). 
In addition, the table also reports the overall MS of the original data in the prediction period, and the in-sample MSE of TFM-cp in the same period, computed using the loading vectors estimated from the full dataset, as references. 
In the last two rows, we report the prediction performance of the first three indices and the others separately. 

From the table, it is seen that T-TFM-cp performed the best considering all indices and indices 4-10 only. The individual ARMA models perform the best for the three CPI indices, 
similar to that has been revealed in Figure~\ref{fig:R2FullData}. It is also evident that there is indeed a threshold phenomenon and using
the threshold model is useful, as seen from the comparison between ARMA-TFM-cp and T-TFM-cp, and the comparison between ARMA-VFM and T-VFM. 

\begin{table}[H]
\centering
{\small
\begin{tabular}{|c|c|c|c|c|c|c|c|}
\hline
Indices & MS($\bX_t$) & TFM-cp & ARMA & VFM & VFM & TFM-cp & TFM-cp \\
& & (MSE) & & ARMA & TAR & ARMA & TAR \\
\hline
\hline
All & 1.1937 & 0.6541 & 0.8494 & 1.0974 & 0.9301 & 1.0222 
& 0.8218
\\
\hline
4-10 & 1.1606 & 0.6183 & 0.9907 & 1.0588 & 0.9264 & 1.0583 
& 0.7735
\\
\hline
1-3 & 1.2710 & 0.7376 & 0.5196 & 1.1873 & 0.9387 & 0.9382 
& 0.9344 \\ 
\hline
\end{tabular}}
\caption{Prediction performance comparison: one-step rolling prediction mean squared error 
of the last 40 observations of five different models. The sample mean squares of observations in the prediction period and 
the in-sample mean square errors of TFM-cp in the prediction period are included as references. 
}
\label{tbl:predError_Split}
\end{table}

\begin{figure}[ht] 
\centering
\includegraphics[width=0.49\linewidth]{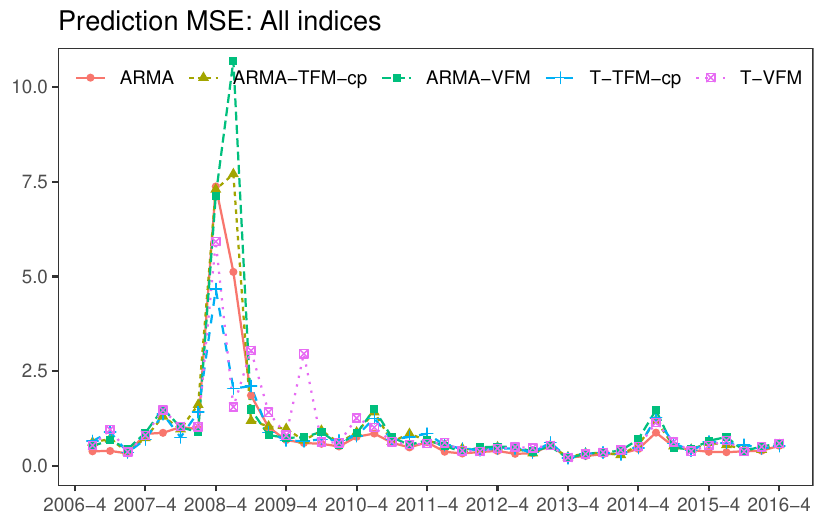}

\includegraphics[width=0.49\linewidth]{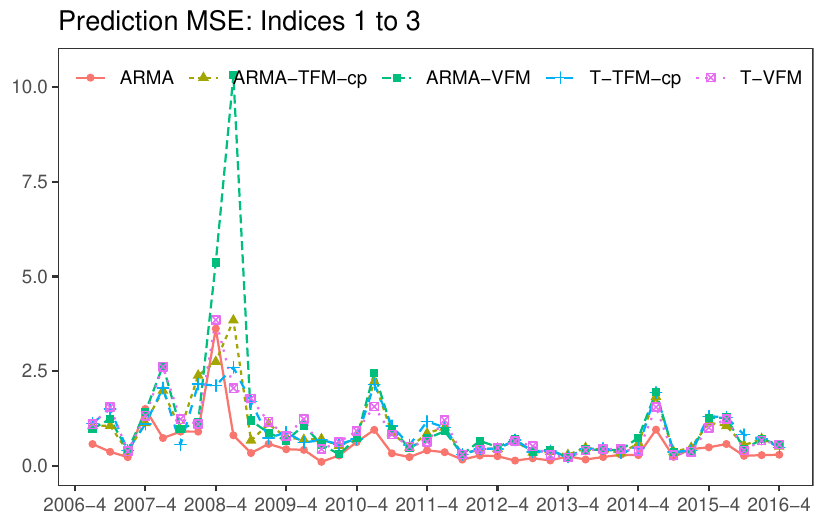}
\includegraphics[width=0.49\linewidth]{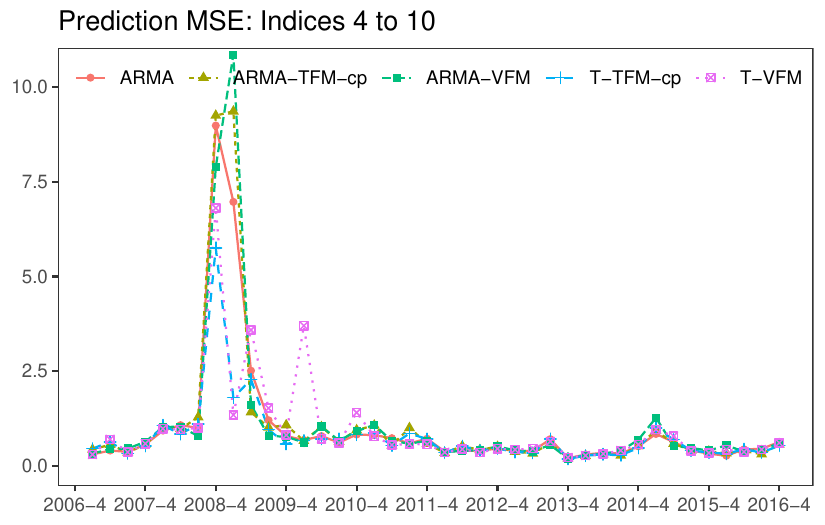}
\caption{Prediction MSE at each $t$ under five different methods.
}\label{PredictionErrorbyTime}
\end{figure}

Figure \ref{PredictionErrorbyTime} shows the prediction MSE of each of the 40 
prediction times. It seems that for the economic indices 4 to 10, the five models are comparable except during the  financial crisis period (mid 2008 to mid 2009), in which T-TFM-cp
out-performs ARMA and ARMA-TFM-cp in 2008, and ARMA-VFM in 2009. For the CPI related indices, individual ARMA models 
perform the best except for the second half of year 2008, 

\section{Conclusions} \label{sec:discussion}

In this paper we proposed a threshold Tensor factor model in CP form, in which the latent factor processes in a TFM-cp model are assumed to follow threshold AR models. It effectively provides 
a threshold type dynamics for high dimensional tensor time series that allows regime change according to the threshold variable. The model also allows prediction capability while enjoys significant dimension reduction using the factor model structure. 

While we have shown that the proposed model is useful in empirical studies, there are challenging issues that are worth further investigation. For example, the two-step estimation procedure works only when the signal to noise ratio in the factor model part is sufficiently high. An effective joint estimation procedure may be needed when the signal to noise ratio is relatively smaller. Hunting for an effective threshold variable is always a challenge in threshold modeling.  





\bibliography{References,mine141104,FRG}

\phantomsection\label{supplementary-material}
\bigskip

\newpage
\begin{center}

{\large\bf Supplementary Material to ``Threshold Tensor Factor Model in CP Form''}

\end{center}

\appendix

In this supplementary material, Appendix \ref{sec:proof} provides the complete proofs of the main theorems presented in the article, while Appendix \ref{A:RealDataFull} presents the estimated models using the full dataset for the real data application.

\section{Proofs of Main Theorems} \label{sec:proof}

\begin{proof}[\bf Proof of Proposition \ref{thm:factors}]
The proof of Proposition \ref{thm:factors} can be derived using similar arguments to those in Theorem 4.3 of \cite{chen2024estimation} and Theorem 3 of \cite{han2024cp}. 
\end{proof}

We first provide a theorem for the consistency of TAR parameters. Given that the least squares estimation of TAR involves a non-convex optimization problem, this result forms the basis for further analysis of the statistical convergence rate and asymptotic normality.
\begin{theorem}\label{thm:consistency}
Suppose Assumptions \ref{asmp:error}, \ref{asmp:eigenvalue}, \ref{asmp:mixing}, \ref{asmp:tar-noise} hold. Assume that $\bphi_j^{(\ell)}\neq \bphi_j^{(\ell+1)}$ for $1\le j\le r,1\le \ell \le L-1$. Let
\begin{align*}
\frac{\sigma \sqrt{d_{\max}} }{\lambda \sqrt{T} }   + \frac{\sigma^2 }{\lambda^2 } \longrightarrow 0,    
\end{align*}
as $T\to \infty$.
\begin{enumerate}
\item[(i)] Suppose $z_{jt}$ is self-exciting with $z_{jt}=f_{j,t-\tau_j}$, for some $1\le j\le r$. Let $\btheta_j=(\bphi_j^{\top},\bs_j^{\top},\tau_j)^{\top}$.
Then, in an event with probability at least $1-T^{-c}-\sum_{k=1}^K e^{-d_k}$, $\widehat \btheta_{j}\to \btheta_j$ as $T\to\infty$.

\item[(ii)] Suppose $z_{jt}$ is an observable variable with a bounded, continuous and positive density on $\R$, and is Markovian, for some $1\le j\le r$. Let $\btheta_j=(\bphi_j^{\top},\bs_j^{\top})^{\top}$. Then, in an event with probability at least $1-T^{-c}-\sum_{k=1}^K e^{-d_k}$, $\widehat \btheta_{j} \to \btheta_j$ as $T\to\infty$.
\end{enumerate}
    
\end{theorem}

\begin{proof}[\bf Proof of Theorem \ref{thm:consistency}]
We only prove part (i). The proof of part (ii) is similar.
Define the squared loss functions with estimated and true factors,
\begin{align} 
\widehat\cL_T(\btheta_j) &=\sum_{t=p_j+1}^T \left(\hat{f}_{j,t}-\sum_{\ell}^L I(z_{j,t}\in(s_{j,\ell-1},s_{j,\ell}))
\sum_{k=1}^{p_j}\phi_{j,k}^{(\ell)}\hat{f}_{j,t-k}\right)^2, \label{LS-TAR:eq1} \\
\cL_T(\btheta_j) &=\sum_{t=p_j+1}^T \left({f}_{j,t}-\sum_{\ell}^L I(z_{j,t}\in(s_{j,\ell-1},s_{j,\ell}))
\sum_{k=1}^{p_j}\phi_{j,k}^{(\ell)} {f}_{j,t-k}\right)^2, \label{LS-TAR:eq2}
\end{align}
and define the residuals
\begin{align}
\widehat \varepsilon_t(\btheta_j) &=\hat{f}_{j,t}-\sum_{\ell}^L I(z_{j,t}\in(s_{j,\ell-1},s_{j,\ell}))
\sum_{k=1}^{p_j}\phi_{j,k}^{(\ell)}\hat{f}_{j,t-k}, \label{res:eq1} \\
\varepsilon_t(\btheta_j) &= {f}_{j,t}-\sum_{\ell}^L I(z_{j,t}\in(s_{j,\ell-1},s_{j,\ell}))
\sum_{k=1}^{p_j}\phi_{j,k}^{(\ell)} {f}_{j,t-k}. \label{res:eq2}
\end{align}
The following two Lemmas can be obtained from \cite{zhang2024least,li2012least}.
\begin{lemma}\label{lemma1:theta}
If the conditions in Theorem \ref{thm:consistency} hold, then $\E\varepsilon_t^2(\widetilde\btheta_j)\ge \E\varepsilon_t^2(\btheta_j)$ for all $\widetilde\btheta_j\in\Theta$, and the equality holds if and only if $\widetilde\btheta_j=\btheta_j$.
\end{lemma}

\begin{lemma}\label{lemma2:theta}
If the conditions of Theorem \ref{thm:consistency} hold, then, for any $\widetilde\btheta_j \in \Theta$ and its open neighborhood $U_{\widetilde\btheta_j}$,
\[
\E \sup_{\btheta_j^* \in U_{\widetilde\btheta_j}} \left| \varepsilon_t^2(\btheta_j^*) - \varepsilon_t^2(\widetilde\btheta_j) \right| \to 0 \quad \text{as } U_{\widetilde\btheta_j} \text{ shrinks to } \widetilde\btheta_j.
\]
\end{lemma}
According to Proposition \ref{thm:factors}, we have in an event $\Omega$ with probability at least $1-T^{-c}-\sum_{k=1}^K e^{-d_k}$, \begin{align}
&\left| \frac{1}{T-h} \sum_{t=h+1}^T \widehat f_{i,t-h} \widehat f_{jt} - \frac{1}{T-h} \sum_{t=h+1}^T f_{i,t-h} f_{jt} \right| \le c \left(  \frac{\sigma \sqrt{d_{\max}} }{\lambda \sqrt{T} }   + \frac{\sigma^2 }{\lambda^2 }  \right)=:\delta.  
\end{align}
In the following proof, we restrict to the event $\Omega$.

For any given open neighborhood $V$ of $\btheta_j \in \Theta$ and any $\vartheta \in V^c = \Theta \backslash V$, $\E \varepsilon_t^2(\vartheta) > \E \varepsilon_t^2(\btheta_j)$ by Lemma \ref{lemma1:theta}. Lemma \ref{lemma2:theta} implies that $\mathbb{E}\varepsilon_t^2(\vartheta)$ is continuous in $\vartheta$. Applying the compactness of $V^c$, there exists a $\vartheta_0 \in V^c$ such that
\[
\inf_{\vartheta \in V^c} \mathbb{E} \varepsilon_t^2(\vartheta) = \mathbb{E} \varepsilon_t^2(\vartheta_0) > \E \varepsilon_t^2(\btheta_j).
\]
Set $\kappa_0 = (\mathbb{E}\varepsilon_t^2(\vartheta_0) - \E \varepsilon_t^2(\btheta_j))/3 > 0$. For any $\vartheta \in V^c$, by Lemma \ref{lemma2:theta} again, there exists $\eta > 0$ such that
\begin{align}\label{eq:theta1}
\mathbb{E} \inf_{\vartheta^* \in U_{\vartheta}(\eta)} \varepsilon_t^2(\vartheta^*) \geq \mathbb{E}\varepsilon_t^2(\vartheta) - \kappa_0 \geq \mathbb{E}\varepsilon_t^2(\vartheta_0) - \kappa_0 = 2\kappa_0 + \E \varepsilon_t^2(\btheta_j).     
\end{align}
Since $V^c$ is compact, there exists a finite covering of $V^c$: $\{U_{\vartheta_j}(\eta), \vartheta_j \in V^c, j = 1, 2, \ldots, J\}$ such that $\bigcup_{j=1}^J U_{\vartheta_j}(\eta) = V^c$. By the ergodic theorem and \eqref{eq:theta1}, we have a.s. for sufficiently large $T$ and $1 \leq j \leq J$
\[
\inf_{\vartheta^* \in U_{\vartheta_j}(\eta)} \frac{1}{T} \widehat\cL_T(\vartheta^*) \geq \frac{1}{T} \sum_{t=1}^T \inf_{\vartheta^* \in U_{\vartheta_j}(\eta)} \hat\varepsilon_t^2(\vartheta^*) \geq \mathbb{E} \inf_{\vartheta^* \in U_{\vartheta_j}(\eta)} \varepsilon_t^2(\vartheta^*) - \kappa_0 -c_1 \delta \geq \kappa_0 + \E \varepsilon_t^2(\btheta_j) - c_1\delta 
\]
and
\[
\inf_{\vartheta \in V} \frac{1}{T} \widehat\cL_T(\vartheta) \leq \frac{1}{T} \widehat\cL_T(\btheta_j) = \frac{1}{T} \sum_{t=1}^T \hat \varepsilon_t^2(\btheta_j) \leq \E \varepsilon_t^2(\btheta_j) + \frac{\kappa_0}{2} + c_1\delta.
\]
Therefore, as $\delta=o(1)$, for any neighborhood $V$ of $\btheta_j \in \Theta$, it follows that for sufficiently large $T$
\[
\inf_{\vartheta^* \in V^c} \frac{1}{T} \widehat\cL_T(\vartheta^*) \geq \min_{1 \leq j \leq J} \inf_{\vartheta^* \in U_{\vartheta_j}(\eta)} \frac{1}{T} \widehat\cL_T(\vartheta^*) \geq \kappa_0 + \E \varepsilon_t^2(\btheta_j)  - c_1 \delta  > \inf_{\vartheta \in V} \frac{1}{T} \widehat\cL_T(\vartheta),
\]
which implies that $\widehat{\btheta}_j \in V$ a.s. By the arbitrariness of $V$, we can get $\widehat{\btheta}_j \to \btheta_j$ in the high probability event $\Omega$. The proof is complete.

\end{proof}

\begin{proof}[\bf Proof of Theorem \ref{thm:sftar}]
According to Proposition \ref{thm:factors}, we have in an event $\Omega$ with probability at least $1-T^{-c}-\sum_{k=1}^K e^{-d_k}$,
\begin{align}
\left|\widehat f_{jt} - f_{jt} \right| \le c \left(\frac{\sigma \sqrt{d_{\max}} }{\lambda \sqrt{T} } + \frac{\sigma}{\lambda} \right) =: \vartheta_1,  
\end{align}
for all $1\le t \le T$, and
\begin{align}
&\left| \frac{1}{T-h} \sum_{t=h+1}^T \widehat f_{i,t-h} \widehat f_{jt} - \frac{1}{T-h} \sum_{t=h+1}^T f_{i,t-h} f_{jt} \right| \le c \left(  \frac{\sigma \sqrt{d_{\max}} }{\lambda \sqrt{T} }   + \frac{\sigma^2 }{\lambda^2 }  \right)=:\vartheta_2,  
\end{align}
for all $1\le i,j\le r$ and $0\le h\le T/4$.
In the following proof, we restrict to the event $\Omega$. For simplicity of presentation, we omit the sub-subscripts and use use $\btheta,\widehat\btheta,\bphi,\bs,\tau,f_{t}$ to represent $\btheta_j,\widehat\btheta_j,\bphi_j,\bs_j,\tau_j,f_{jt}$, etc. Without loss of generality, consider two regime TAR models and let $L=2$. The extension to general $L$ follows similarly.

By Theorem \ref{thm:consistency}, in the event $\Omega$, $\widehat\btheta\to\btheta$. Thus, in $\Omega$, $\hat\tau_j=\tau_j$, and we restrict to the parameter space of an open neighborhood of $\btheta$. Define $V_{\delta}=\{\widetilde\btheta\in\Theta: \|\widetilde\bphi-\bphi\|<\delta, |\widetilde\bs-\bs|< \delta\} $ for some $0<\delta<1$ to be determined later.

Define the residuals
\begin{align}
\widehat \varepsilon_t(\bphi_0,\bs_0) &=\hat{f}_{t}-\sum_{\ell=1}^2 I(\hat f_{t-\tau}\in(s_{0,\ell-1},s_{0,\ell}))
\sum_{k=1}^{p_j}\phi_{0k}^{(\ell)}\hat{f}_{t-k},  \\
\varepsilon_t(\bphi_0,\bs_0) &= {f}_{t}-\sum_{\ell=1}^2 I(f_{t-\tau}\in(s_{0,\ell-1},s_{0,\ell}))
\sum_{k=1}^{p_j}\phi_{0k}^{(\ell)} {f}_{t-k}, 
\end{align}
with $s_{0,0}=-\infty, s_{0,2}=+\infty$,
and the squared loss functions with estimated and true factors,
\begin{align} 
\widehat\cL_T(\bphi_0,\bs_0) &=\sum_{t=p_j+1}^T \widehat \varepsilon_t^2(\bphi_0,\bs_0), \label{LS-TAR:eq3} \\
\cL_T(\bphi_0,\bs_0) &=\sum_{t=p_j+1}^T \varepsilon_t^2 (\bphi_0,\bs_0). \label{LS-TAR:eq4}
\end{align}
Note that the difference of $\widehat\varepsilon_t(\bphi_0,\bs_0)$ and $\widehat\varepsilon_t(\bphi_0,\bs)$ is nonzero only in the region $\mathbf{1}\{s_0 \wedge s < \widehat f_{t-\tau} \le s_0\vee s\}$. Here we only treat the case $s_0>s$. Proofs of the other case is similar. Write $s_0=s+u$ for some $0<u<1$, and let $\widehat\alpha_{t}^{(\ell)}= (\hat f_t -\bphi^{(\ell)\top} \hat\bff_{t-1})^2$, $\widehat\alpha_{t}^{(\ell0)}= (\hat f_t -\bphi_0^{(\ell)\top} \hat\bff_{t-1})^2$ and $\alpha_{t}^{(\ell)}= ( f_t -\bphi^{(\ell)\top} \bff_{t-1})^2$, $\alpha_{t}^{(\ell0)}= ( f_t -\bphi_0^{(\ell)\top} \bff_{t-1})^2$, where $\hat\bff_{t-1} =(\hat f_{t-1},...,\hat f_{t-p_j})^\top$.
Then
\begin{align*}
R_T(\bphi_0,u)&=  \widehat\cL_T(\bphi_0,s_0)  - \widehat\cL_T(\bphi_0,s)
=\sum_{t=p_j+1}^T (\widehat \alpha_t^{(20)} - \widehat \alpha_t^{(10)} ) \mathbf{1}\{ s< \hat f_{t-\tau} \le s+u\} \\
&= [\widehat\cL_T(\bphi,s_0)  - \widehat\cL_T(\bphi,s)] + \{ [\widehat\cL_T(\bphi_0,s_0)  - \widehat\cL_T(\bphi,s_0)] - [\widehat\cL_T(\bphi_0,s)  - \widehat\cL_T(\bphi,s)]\} \\
&:= R_{T1}(u) +R_{T2}(\bphi_0,u) ,
\end{align*}
where
\begin{align*}
R_{T1}(u) &=  \sum_{t=p_j+1}^T (\widehat \alpha_t^{(2)} - \widehat \alpha_t^{(1)} ) \mathbf{1}\{ s< \hat f_{t-\tau} \le s+u\} , \\
R_{T2}(\bphi_0,u)   &= \sum_{t=p_j+1}^T [(\widehat \alpha_t^{(20)} - \widehat \alpha_t^{(2)} ) - (\widehat \alpha_t^{(10)} - \widehat \alpha_t^{(1)} )]  \mathbf{1}\{ s< \hat f_{t-\tau} \le s+u\}.
\end{align*}

For $R_{T1}(u)$, by Assumption \ref{asmp:tar-iden}, for all $i\neq j$, there exist some positive constants $c_0$ and $\rho$ such that $|(\bphi^{(i)}- \bphi^{(j)})^\top \bff_{t-1}|\ge c_0>0$ for all $\| \bff_{t-1} -\bw_1\| \le \rho$. Then
\begin{align}\label{tareq1}
[(\bphi^{(i)}- \bphi^{(j)})^\top \bff_{t-1}]^2 + 2\xi_t (\bphi^{(i)}- \bphi^{(j)})^\top \bff_{t-1} \ge c_0^2 \mathbf{1}\{ \| \bff_{t-1} -\bw_1\| \le \rho\}   +  2\xi_t (\bphi^{(i)}- \bphi^{(j)})^\top \bff_{t-1} .
\end{align}
Let $\varpi:=\max_{i\neq j} \| \bphi^{(i)} - \bphi^{(j)}\|.$ By \eqref{tareq1} and Lemma \ref{lemma:tar}, for $R_{T1}(u)$, we have
\begin{align*}
R_{T1}(u) \ge& c_0^2 G_T^*(u-2\vartheta_1) -2 \varpi \left( \left| \sum_{t=p_j+1}^T A_t(u+2\vartheta_1) \right| + \left| \sum_{t=p_j+1}^T D_t(u+2\vartheta_1) \right| \right) \\
&- 3(\varpi^2+2\varpi) \vartheta_1 G_T(u+2\vartheta_1),   
\end{align*}
where $G_T^*(u-2\vartheta_1), A_t(u+2\vartheta_1), D_t(u+2\vartheta_1)$, and other notations used below are defined in Lemma \ref{lemma:tar}. Thus
\begin{align*}
& \inf_{\frac{B}{T}<u<\delta} \frac{R_{T1}(u)}{T G(u-2\vartheta_1)}    \\
\ge& c_0^2  \inf_{\frac{B}{T}<u<\delta} \frac{G_T^*(u-2\vartheta_1)}{T G^*(u-2\vartheta_1)} \frac{G^*(u-2\vartheta_1)}{G(u-2\vartheta_1)} \\
&\quad -2\varpi 
\sup_{\frac{B}{T}<u<\delta} \frac{  \left| \sum_{t=p_j+1}^T A_t(u+2\vartheta_1) \right| + \left| \sum_{t=p_j+1}^T D_t(u+2\vartheta_1) \right| }{T G(u+2\vartheta_1) } \frac{G(u+2\vartheta_1)} {G(u-2\vartheta_1)} \\
&\quad - \inf_{\frac{B}{T}<u<\delta}\frac{3(\varpi^2+2\varpi) \vartheta_1 G_T(u+2\vartheta_1)}{T G(u-2\vartheta_1)} \\
\ge&  c_0^2  \left( 1- \sup_{\frac{B}{T}<u<\delta} \left| \frac{G_T^*(u-2\vartheta_1)}{T G^*(u-2\vartheta_1)} -1  \right|   \right)  \frac{G^*(u-2\vartheta_1)}{G(u-2\vartheta_1)} - \inf_{\frac{B}{T}<u<\delta}\frac{3(\varpi^2+2\varpi) \vartheta_1 G_T(u+2\vartheta_1)}{T G(u-2\vartheta_1)} \\
&\quad -2\varpi 
\sup_{\frac{B}{T}<u<\delta} \frac{  \left| \sum_{t=p_j+1}^T A_t(u+2\vartheta_1) \right| + \left| \sum_{t=p_j+1}^T D_t(u+2\vartheta_1) \right| }{T G(u+2\vartheta_1) } \frac{G(u+2\vartheta_1)} {G(u-2\vartheta_1)}  .
\end{align*}
Note that as $\vartheta_1=o(1)$,
\begin{align*}
 \frac{G^*(u-2\vartheta_1)}{G(u-2\vartheta_1)}   &=\P (\|\bff_{t-1}-\bw_1\|\le \rho | s < f_{t-\tau}\le s+u-2\vartheta_1) \\
&\longrightarrow \P (\|\bff_{t-1}-\bw_1\|\le \rho |  f_{t-\tau}=s) >0 \quad \text{as } u-2\vartheta_1\downarrow 0 ,
\end{align*}
which implies that the infimum below exists for sufficiently small $\delta>2\vartheta_1>0$, namely,
\begin{align*}
c_1:=  \inf_{0<u<\delta}\frac{G^*(u-2\vartheta_1)}{G(u-2\vartheta_1)} >0.
\end{align*}
As $\vartheta_1,\vartheta_2\to0$, choose $\eta_0>0$ such that $2\gamma=c_0^2 c_1 -(c_0^2+8\varpi)\eta_0-6(\varpi^2+2\varpi) \vartheta_1 \eta_0$. By Lemma \ref{lemma:tar}, it follows that
\begin{align}\label{tareq2}
&\P\left(  \inf_{\frac{B}{T}<u<\delta} \frac{R_{T1}(u)}{T G(u-2\vartheta_1)}  > 2\gamma \right) \notag\\
\ge& \P\left( \sup_{\frac{B}{T}<u<\delta} \left| \frac{G_T^*(u-2\vartheta_1)}{T G^*(u-2\vartheta_1)} -1  \right|<\eta_0,  \sup_{\frac{B}{T}<u<\delta} \frac{  \left| \sum_{t=p_j+1}^T A_t(u+2\vartheta_1) \right|  }{T G(u+2\vartheta_1) }< \eta_0 , \right. \notag\\
&\quad\left. \sup_{\frac{B}{T}<u<\delta} \left| \frac{G_T(u-2\vartheta_1)}{T G(u-2\vartheta_1)} -1  \right|<\eta_0,  \sup_{\frac{B}{T}<u<\delta} \frac{   \left| \sum_{t=p_j+1}^T D_t(u+2\vartheta_1) \right| }{T G(u+2\vartheta_1) }< \eta_0 \right) \notag \\
\ge& 1- \varepsilon.
\end{align}
Note that
\begin{align*}
&[(\widehat \alpha_t^{(20)} - \widehat \alpha_t^{(2)} ) - (\widehat \alpha_t^{(10)} - \widehat \alpha_t^{(1)} )]  \mathbf{1}\{ s< \hat f_{t-\tau} \le s+u\} \notag\\
&= \big\{[(\bphi^{(2)}- \bphi^{(20)})^\top \hat\bff_{t-1}]^2 - [(\bphi^{(1)}- \bphi^{(10)})^\top \hat\bff_{t-1}]^2 + (\bphi^{(2)}- \bphi^{(20)})^\top \hat\bff_{t-1} (\bphi^{(1)}- \bphi^{(10)})^\top \hat\bff_{t-1}\\
&\qquad + 2\hat \xi_t  (\bphi^{(10)}- \bphi^{(1)}+\bphi^{(2)}- \bphi^{(20)})^\top \hat\bff_{t-1} 
\end{align*}

For $R_{T2}(\bphi_0,u)$, by the above identity and Lemma \ref{lemma:tar}, we have
\begin{align*}
|R_{T2}(\bphi_0,u)|   \le& C\delta \left((\E\|\bff_{t-1}\|^2 + \vartheta_2) G_T(u+2\vartheta_1) +\left| \sum_{t=p_j+1}^T H_t(u+2\vartheta_1) \right| +\left| \sum_{t=p_j+1}^T A_t(u+2\vartheta_1) \right| \right. \\
&\quad \left.+ \left| \sum_{t=p_j+1}^T D_t(u+2\vartheta_1) \right| \right).
\end{align*}
It implied that
\begin{align}\label{tareq3}
&\sup_{\frac{B}{T}<u<\delta,\btheta\in V_{\delta}} \frac{|R_{T2}(\bphi_0,u)|}{TG(u-2\vartheta_1)} \notag\\
&\le    C\delta \left((\E\|\bff_{t-1}\|^2 + \vartheta_2) \left[ \sup_{\frac{B}{T}<u<\delta} \left| \frac{G_T(u+2\vartheta_1)}{T G(u-2\vartheta_1)} -1  \right| +1 \right]+ \right. \notag\\
&\quad \left.+ \sup_{\frac{B}{T}<u<\delta}\frac{\left| \sum_{t=p_j+1}^T H_t(u+2\vartheta_1) \right| +\left| \sum_{t=p_j+1}^T A_t(u+2\vartheta_1) \right| + \left| \sum_{t=p_j+1}^T D_t(u+2\vartheta_1) \right|}{TG(u-2\vartheta_1)} \right) \notag\\
&=O_{\P}(\delta).
\end{align}
Then by \eqref{tareq2} and \eqref{tareq3}, for sufficiently small $\delta>0$, we have
\begin{align*}
&\P\left(  \inf_{\frac{B}{T}<u<\delta} \frac{R_{T}(u)}{T G(u-2\vartheta_1)}  > \gamma \right)   \ge \P\left(  \inf_{\frac{B}{T}<u<\delta} \frac{R_{T1}(u)}{T G(u-2\vartheta_1)} -  \sup_{\frac{B}{T}<u<\delta,\btheta\in V_{\delta}} \frac{|R_{T2}(\bphi_0,u)|}{TG(u-2\vartheta_1)}  > \gamma \right) \\
\ge& 1 - \P\left(  \inf_{\frac{B}{T}<u<\delta} \frac{R_{T1}(u)}{T G(u-2\vartheta_1)} \le 2 \gamma \right)  -\P\left(   \sup_{\frac{B}{T}<u<\delta,\btheta\in V_{\delta}} \frac{|R_{T2}(\bphi_0,u)|}{TG(u-2\vartheta_1)} \ge \gamma \right) \\
\ge& 1-2\epsilon.
\end{align*}
Thus, we finish the proof of \eqref{thm_sftar_eq1}.

Next, we move to the proof of \eqref{thm_sftar_eq2} and \eqref{thm_sftar_eq3}.

Let \(\widehat\ell_T(\bphi_0,\bs_0)=\widehat\cL_T(\bphi_0,\bs_0)/T\) and \(\ell_T(\bphi_0,\bs_0)=\cL_T(\bphi_0,\bs_0)/T\). By the Taylor expansion of \(\partial \widehat\ell_T(\bphi_0,\bs_0)/\partial\bphi_0\), we have  
\begin{align}\label{tarclt:eq1}
0=\frac{\partial \widehat\ell_T(\widehat\bphi(\bs_0),\bs_0)}{\partial\bphi_0}
=\frac{\partial \widehat\ell_T(\bphi,\bs_0)}{\partial\bphi_0}
+\frac{\partial^2 \widehat\ell_T(\widetilde\bphi,\bs_0)}{\partial\bphi_0\partial\bphi_0'}
 \bigl(\widehat\bphi(\bs_0)-\bphi\bigr),    
\end{align}
where \(\widetilde\bphi\) lies in between \(\widehat\bphi(\bs_0)\) and \(\bphi\), i.e.,
\(\|\widetilde\bphi-\bphi\|\le\|\widehat\bphi(\bs_0)-\bphi\|\).

Let \(\widetilde\Sigma=\mathrm{diag}(\Sigma_1,\Sigma_2)\). Since \(\E(\hat f_t^2)<\infty\) and $\vartheta_1=o(1),\vartheta_2=o(1)$, by the law of large numbers, it follows that  
\[
\frac{\partial^2 \widehat\ell_T(\btheta)}{\partial\bphi_0\partial\bphi_0'}
\;\to\;2\widetilde\Sigma,
\qquad\text{a.s. as } T\to\infty.
\]
Then, by \eqref{tarclt:eq1} and Lemma \ref{lemma:likelihood}, we have
\[
\sup_{\|\bs - \bs_0\|\le B(1/T +\vartheta_1)}
\left\|
\bigl[\widehat\bphi(\bs_0)-\bphi\bigr]
+(2\widetilde\Sigma)^{-1}\frac{\partial \ell_T(\bphi,\bs)}{\partial\bphi_0}
\right\|
=O_{\P}\left( \frac{1}{\sqrt{T}} + \vartheta_1 + \vartheta_2  \right).
\]
This implies that \eqref{thm_sftar_eq2}.

Furthermore, if $\lambda/\sigma \gg T^{1/2}+ \sqrt{d_{\max}}$, then $\vartheta_1+\vartheta_2=o(T^{-1/2})$. It follows that
\begin{align*}
\sup_{\|\bs - \bs_0\|\le B(1/T +\vartheta_1)}
\sqrt{T} \|\widehat\bphi(\bs_0)-\widehat\bphi(\bs)\|
\le&
\sup_{\|\bs - \bs_0\|\le B(1/T +\vartheta_1)}
\left\|
\sqrt{T}\bigl[\widehat\bphi(\bs_0)-\bphi\bigr]
+(2\widetilde\Sigma)^{-1}\sqrt{T}\frac{\partial\ell_T(\bphi,\bs)}{\partial\bphi_0}
\right\| \\
&
+\left\|
\sqrt{T}\bigl[\widehat\bphi(\bs)-\bphi\bigr]
+(2\widetilde\Sigma)^{-1}\sqrt{T}\frac{\partial\ell_T(\bphi,\bs)}{\partial\bphi_0}
\right\|
=o_{\P}(1).  
\end{align*}
Note that
\[
\sqrt{T}\frac{\partial\ell_T(\bphi,\bs)}{\partial\bphi_0}
=\frac{2}{\sqrt{T}}
\sum_{t=p_j+1}^T
\frac{\partial \varepsilon_t(\bphi,\bs)}{\partial\bphi_0} \varepsilon_t(\bphi,\bs),
\]
and \(\{\varepsilon_t(\bphi,\bs)\,\partial \varepsilon_t(\bphi,\bs)/\partial\bphi_0\}\) is a martingale difference sequence
with respect to \(\{\mathcal{F}_t\}\).
By the martingale central limit theorem, it follows that
\[
\sqrt{T}\frac{\partial\ell_T(\bphi,\bs)}{\partial\bphi_0}
\;\xrightarrow{d}\;
\mathcal{N}(0,4 \bar\Sigma),
\]
where
\[
\bar\Sigma=\mathrm{diag}  ((\varsigma^{(1)})^2\Sigma_1,\;(\varsigma^{(2)})^2\Sigma_2).
\]
Thus,
\[
\sqrt{T}\bigl(\widehat\bphi-\bphi\bigr)
\xrightarrow{d}
\mathcal{N}(0,\bSigma).
\]
The proof of \eqref{thm_sftar_eq3} is complete.

\end{proof}

\begin{proof}[\bf Proof of Theorem \ref{thm:obtar}]
The proof of Theorem \ref{thm:obtar} closely parallels that of Theorem \ref{thm:sftar}, with the key distinction that the threshold variable is observable. Consequently, we can eliminate the term $\vartheta_1$ that appears in the original proof of Theorem \ref{thm:sftar}, thereby obtaining improved convergence rates. The details are omitted due to their similarity.
\end{proof}

\begin{lemma}\label{lemma:tar}
If Assumptions \ref{asmp:mixing}, \ref{asmp:tar-noise}, \ref{asmp:tar-trans}, \ref{asmp:tar-iden} hold, then, for any $\epsilon > 0$, $\eta > 0$ and $\delta \in (0,1)$, there exists a positive constant $B$ such that for $T$ large enough, we have

\begin{enumerate}[label=(\roman*).]
\item $\mathbb{P}\left( \sup_{B/T < u < \delta} \left| \frac{G_T(u)}{TG(u)} - 1 \right| < \eta \right) > 1 - \epsilon$,

\item $\mathbb{P}\left( \sup_{B/T < u < \delta} \left| \frac{G_T^*(u)}{TG^*(u)} - 1 \right| < \eta \right) > 1 - \epsilon$,

\item $\mathbb{P}\left( \sup_{B/T < u < \delta} \frac{\left| \sum_{t=1}^T A_t(u) \right|}{TG(u)} < \eta \right) > 1 - \epsilon$,

\item $\mathbb{P}\left( \sup_{B/T < u < \delta} \frac{\left| \sum_{t=1}^T D_t(u) \right|}{TG(u)} < \eta \right) > 1 - \epsilon$,

\item $\mathbb{P}\left( \sup_{B/T < u < \delta} \frac{\left| \sum_{t=1}^T H_t(u) \right|}{TG(u)} < \eta \right) > 1 - \epsilon$,
\end{enumerate}
where
\begin{align*}
G(u) &= \mathbb{P}(s < z_{t} \leq s + u), \\
G_T(u) &= \sum_{t=1}^T I(s < z_{t} \leq s + u), \\
G_T^*(u) &= \sum_{t=1}^T I(s < z_{t} \leq s + u, \|\bff_{t-1} - \bw_1\| \leq \rho ), \\
G^*(u) &= \mathbb{P}(s < z_{t} \leq s + u, \|\bff_{t-1} - \bw_1\| \leq \rho ), \\
A_t(u) &= \xi_t I(s < z_{t} \leq s + u), \\
D_t(u) &= \bff_{t-1} \xi_t I(s < z_{t} \leq s + u), \\
H_t(u) &= (\|\bff_{t-1}\|^2 - \mathbb{E}\|\bff_{t-1}\|^2)I(s < z_{t} \leq s + u).
\end{align*}
\end{lemma}

\begin{lemma}\label{lemma:likelihood}
If the conditions in Theorem \ref{thm:sftar} hold, then, for any \(0 < B < \infty\),
\begin{align}
\sup_{\|\bs - \bs_0\| \le B(1/T +\vartheta_1)}
\left\|
\frac{\partial \widehat\ell_T(\bphi,\bs_0)}{\partial \bphi_0}
-
\frac{\partial \widehat\ell_T(\btheta)}{\partial \bphi_0}
\right\|
&= O_{\P}(T^{-1}+\vartheta_1),    \\
\sup_{\|\bs - \bs_0\| \le B(1/T +\vartheta_1)}
\left\|
\frac{\partial^2 \widehat\ell_T(\bphi_0,\bs_0)}{\partial \bphi_0 \partial \bphi_0'}
-
\frac{\partial^2 \widehat\ell_T(\btheta)}{\partial \bphi_0 \partial \bphi_0'}
\right\|
&= O_{\P}(T^{-1}+\vartheta_1),
\end{align}
where
\begin{align*}
\vartheta_1=c \left(\frac{\sigma \sqrt{d_{\max}} }{\lambda \sqrt{T} } + \frac{\sigma}{\lambda} \right) .      
\end{align*}

\end{lemma}

\begin{lemma}\label{lemma:likelihood2}
If the conditions in Theorem \ref{thm:obtar} hold, then, for any \(0 < B < \infty\),
\begin{align}
\sup_{\|\bs - \bs_0\| \le B/T}
\left\|
\frac{\partial \widehat\ell_T(\bphi,\bs_0)}{\partial \bphi_0}
-
\frac{\partial \widehat\ell_T(\btheta)}{\partial \bphi_0}
\right\|
&= O_{\P}(T^{-1}),    \\
\sup_{\|\bs - \bs_0\| \le B/T}
\left\|
\frac{\partial^2 \widehat\ell_T(\bphi_0,\bs_0)}{\partial \bphi_0 \partial \bphi_0'}
-
\frac{\partial^2 \widehat\ell_T(\btheta)}{\partial \bphi_0 \partial \bphi_0'}
\right\|
&= O_{\P}(T^{-1}).
\end{align}

\end{lemma}

Lemma \ref{lemma:tar} is obtained directly from \cite{zhang2024least}, while Lemmas \ref{lemma:likelihood} and \ref{lemma:likelihood2} can be derived using similar arguments as in \cite{zhang2024least}.
    \section{Additional Real Data Application Results}

Table \ref{tbl:models105} shows the estimated models using the full data set. Note that the residual variance and AIC values are all significantly higher than that in Table~\ref{tbl:models65}. This is due to the fact that the full data include the priod of financial crises, with large deviations from the majority of the data.

\begin{table}[H]
\centering
\resizebox{\textwidth}{!}{%
\begin{tabular}{|p{1.6cm}|c|l|l|c|c|c|}
\hline
\textbf{Model} & \textbf{Series} & \textbf{Regime} & \textbf{Coefficients} & \textbf{Threshold} & $\sigma^2$ & \textbf{AIC} \\ \hline
\multirow{2}{*}[-0.5em]{\parbox{1.8cm}{\centering \textbf{TFM-cp ARMA}}}
 & factor1 & -- &
 \begin{tabular}[c]{@{}l@{}}ar1: 0.70 (0.07)\end{tabular}
 & -- & 7.807 & 517.42 \\ \cline{2-7}
 & factor2 & -- &
 \begin{tabular}[c]{@{}l@{}}ma1: 0.98 (0.09)\\ma2: 0.40 (0.09)\end{tabular}
 & -- & 5.993 & 490.99 \\ \hline
\multirow{2}{*}[-0.5em]{\parbox{1.6cm}{\centering \textbf{VFM ARMA}}}
 & factor1 & -- &
 \begin{tabular}[c]{@{}l@{}}ma1: 1.00 (0.10)\\ma2: 0.40 (0.10)\end{tabular}
 & -- & 17.47 & 603.36 \\ \cline{2-7}
 & factor2 & -- &
 \begin{tabular}[c]{@{}l@{}}ar1: 0.62 (0.09)\\ma1: 0.42 (0.11)\end{tabular}
 & -- & 4.901 & 469.99 \\ \hline
\multirow[c]{4}{*}[-2em]{\parbox{1.6cm}{\centering \textbf{TFM-cp TAR}}}
 & \multirow{2}{*}{factor1} & \textbf{Regime 1} (35 obs) &
 \begin{tabular}[c]{@{}l@{}}ar1: 1.21 (0.13)\\ar2: -0.35 (0.12)\end{tabular}
 & \multirow{2}{*}{\parbox{1.5cm}{\centering 0.0103 ($d=4$)}} 
 & \multirow{2}{*}{5.7616}
 & \multirow{2}{*}{179.76} \\ \cline{3-4}
 &  & \textbf{Regime 2} (66 obs) &
 \begin{tabular}[c]{@{}l@{}}ar1: 0.44 (0.09)\end{tabular}
 &  &  &  \\ 
 \cline{2-7}
 & \multirow{2}{*}{factor2} & \textbf{Regime 1} (49 obs) &
 \begin{tabular}[c]{@{}l@{}}ar1: 1.03 (0.14)\\ar2: -0.76 (0.18)\\ar3: 0.45 (0.18)\\ar4: -0.31 (0.13)\end{tabular}
 & \multirow{2}{*}{\parbox{1.5cm}{\centering 0.0118 ($d=4$)}}
 & \multirow{2}{*}{5.1425}
 & \multirow{2}{*}{169.24} \\ \cline{3-4}
 &  & \textbf{Regime 2} (52 obs) &
 \begin{tabular}[c]{@{}l@{}}ar1: 0.78 (0.09)\end{tabular}
 &  &  &  \\ \hline
\multirow{4}{*}[-0.5em]{\parbox{1.6cm}{\centering \textbf{VFM TAR}}}
 & \multirow{2}{*}{factor1} & \textbf{Regime 1} (50 obs) &
 \begin{tabular}[c]{@{}l@{}}ar1: 1.08 (0.14)\\ar2: -0.94 (0.21)\\ar3: 0.54 (0.20)\\ar4: -0.36 (0.14)\end{tabular}
 & \multirow{2}{*}{\parbox{1.5cm}{\centering 0.0119 ($d=4$)}}
 & \multirow{2}{*}{15.6982}
 & \multirow{2}{*}{261.11} \\ \cline{3-4}
 &  & \textbf{Regime 2} (51 obs) &
 \begin{tabular}[c]{@{}l@{}}ar1: 0.73 (0.09)\end{tabular}
 &  &  &  \\ \cline{2-7}
 & \multirow{2}{*}{factor2} & \textbf{Regime 1} (42 obs) &
 \begin{tabular}[c]{@{}l@{}}ar1: 1.13 (0.13)\\ar2: -0.49 (0.13)\end{tabular}
 & \multirow{2}{*}{\parbox{1.5cm}{\centering 0.0112 ($d=2$)}}
 & \multirow{2}{*}{4.2950}
 & \multirow{2}{*}{149.37} \\ \cline{3-4}
 &  & \textbf{Regime 2} (61 obs) &
 \begin{tabular}[c]{@{}l@{}}ar1: 0.65 (0.09)\end{tabular}
 &  &  &  \\ \hline
\end{tabular}
}
\caption{Estimated parameters of ARMA and TAR models fitted to the factor process $\widehat{f}_{1t}$, based on the entire sample (indices 1–10). 
}
\label{tbl:models105}
\end{table}


\label{A:RealDataFull}

\end{document}